\documentclass[aps,twocolumn,nofootinbib]{revtex4-2}

\setcitestyle{authoryear,round}

\usepackage{microtype}
\usepackage{graphicx}
\usepackage{xspace}
\usepackage{hyperref}

\makeatletter
\renewcommand\subsubsection{%
  \@startsection{subsubsection}{3}{0pt}%
  {1.25ex plus .2ex minus .2ex}
  {0.8ex plus .2ex}
  {\normalfont\normalsize\bfseries
   \@afterindentfalse\@afterheading}}
\makeatother

\makeatletter
\renewcommand\subsection{%
  \@startsection{subsection}{2}{0pt}%
  {1.75ex plus .3ex minus .2ex}
  {.9ex plus .2ex}
  {\normalfont\normalsize\bfseries\raggedright
   \@afterindentfalse\@afterheading}}
\makeatother

\newcommand{\micron}{\textmu m\xspace}
\newcommand{\gcm}{g\,cm$^{-3}$\xspace}
\newcommand{\degC}{$^\circ$C\xspace}

\makeatletter

\renewcommand{\@seccntformat}[1]{%
  \csname the#1\endcsname\enspace}

\AtBeginDocument{%
  \let\RT@oldrefstepcounter\refstepcounter
  \renewcommand{\refstepcounter}[1]{%
    \RT@oldrefstepcounter{#1}%
    \protected@edef\@currentlabel{\csname the#1\endcsname}}%
}

\makeatother
\makeatletter
\renewcommand{\p@section}{}
\renewcommand{\p@subsection}{}
\renewcommand{\p@subsubsection}{}
\makeatother

\begin{document}

\title{The science from asteroid sample return missions}

\author{E.~J. Tasker}
\email{elizabeth.tasker@jaxa.jp}
\affiliation{Institute of Space and Astronautical Science,
Japan Aerospace Exploration Agency,
Yoshinodai 3-1-1, Sagamihara, Kanagawa 252-5210, Japan}

\author{H.~C. Connolly Jr}
\affiliation{Department of Geology, Rowan University,
Glassboro, New Jersey, USA}
\affiliation{Lunar and Planetary Laboratory,
University of Arizona, Tucson, Arizona, USA}
\affiliation{Department of Earth and Planetary Sciences,
American Museum of Natural History,
New York, New York, USA}

\author{S. Tachibana}
\affiliation{Institute of Space and Astronautical Science,
Japan Aerospace Exploration Agency,
Yoshinodai 3-1-1, Sagamihara, Kanagawa 252-5210, Japan}
\affiliation{Department of Earth and Planetary Science,
The University of Tokyo, Japan}

\begin{abstract}
\noindent\leaders\hrule\hfill\kern0pt\\
\vspace{0.6ex}
\normalsize
To date, three samples from near-Earth asteroids have been delivered to Earth by
Japan's \textit{Hayabusa} (2010) and \textit{Hayabusa2} (2020) missions, and the
United States \textit{OSIRIS-REx} mission (2023).
Free from terrestrial contamination, these pristine materials provide new
opportunities to investigate planetary formation processes, the delivery of
organics and water to the early Earth, and the nature of potentially hazardous
asteroids. As analysis of the asteroid samples proceeds in laboratories around
the world, we visit each of the missions, review the initial scientific findings,
and explore the value of sample return in understanding our origins and
protecting our future.\\
\vspace{0.6ex}
\noindent\leaders\hrule\hfill\kern10pt
\end{abstract}


\maketitle

\section{Introduction}
On 13 June 2010, the reentry capsule for the JAXA (Japan Aerospace Exploration Agency) Hayabusa mission touched down in the Australian desert. Inside the protective shell was a sample container holding material gathered from asteroid Itokawa. It had been a harrowing journey, but the first asteroid sample return mission had been a success. In December 2020, the JAXA Hayabusa2 mission returned a second sample from asteroid Ryugu, followed in September 2023 with a sample delivered by the NASA Origins, Spectral Interpretation, Resource Identification, and Security-Regolith Explorer (OSIRIS-REx) mission from asteroid Bennu. With these missions, pristine material from the small bodies of our Solar System could now be studied without interference from our atmosphere or terrestrial contamination, ushering in a new era of planetary science.

Asteroid sample return missions represent some of the most technically demanding robotic exploration that has been attempted by humanity. The collection of a sample from an asteroid requires a complex series of autonomous operations to undertake the landing, acquisition, and take-off that must be designed in the absence of knowledge of the asteroid surface properties, which cannot be resolved prior to the spacecraft arrival. And while most robotic exploration involves a one-way journey to a target destination, a sample return mission must also travel back to Earth, significantly increasing the required mission duration. The sample must then be protected within a reentry capsule while shock-heated temperatures between the capsule and atmosphere reach thousands of degrees before the planet surface is reached. Strict protocols have to be implemented to keep the sample free from contamination from the Earth's environment that begin during the construction of the spacecraft and continue throughout the sample collection and analysis. If any step fails, the sample may be compromised or even lost. 

Despite the challenges that such missions present, the scientific incentive for collecting an asteroid sample is high. A central objective has been to better understand the formation and early evolution of terrestrial planets, including the emergence of an environment conducive to habitability and life on Earth. Much of the evidence of conditions on the newly formed Earth has been erased by billions of years of geological, chemical, and biological activity. In contrast, less massive celestial bodies such as asteroids, comets, and small moons have undergone far less geologic evolution. This small body population therefore preserves primitive material that formed near the beginning of the Solar System approximately 4.56 billion years ago, and which has undergone relatively few changes since that epoch. Samples from this population are a chance to examine the building blocks of the early Earth, providing evidence of the starting conditions from which life can develop. 

A second increasingly important goal is to understand the physical and dynamical properties of the small bodies themselves. Historical celestial impact events such as the Tunguska Event in 1908 and the Chelyabinsk meteor in 2013, have demonstrated that collisions with asteroids just tens of meters in diameter can cause wide-spread damage. The prediction of such events, and the development of strategies to prevent a potential collision, require knowledge of the asteroid characteristics. The future trajectory of an asteroid is determined not only by gravitational interactions and minor collisions, but also by thermal forces. These include the Yarkovsky and Yarkovsky–O’Keefe–Radzievskii–Paddack (YORP) effects, in which the momentum carried by photons of solar radiation imparts a miniscule but cumulative force that can significantly alter the orbit and spin of an asteroid over time. The magnitude of these changes depends on properties such as the shape and thermal inertia of an asteroid, making it challenging to accurately predict when an asteroid might hit the Earth without direct data. This was the case with the Chelyabinsk meteor, during whose arrival most astronomers were (rather ironically) watching the successfully predicted close approach of asteroid Duende. Should it become necessary to deflect an asteroid, the effectiveness of any strategy will depend on factors such as the asteroid's composition and tensile strength. Investigating the properties of different classes of asteroid through in-situ exploration and sample return has therefore become an important part of planetary defence research.

Our knowledge of asteroids has been shaped chiefly by ground-based observations and the study of meteorites. These approaches have provided valuable insights into asteroid composition, dynamics, and origins, and have established the framework for our current understanding of the asteroid population. However, there are a number of inherent limitations to studying these small bodies from Earth. Before leaving our planet with the three sample return missions, it is important to consider what has been revealed by Earth-based techniques, and where these fall short. 

\subsection{Meteorites}
\label{sec:meteorites}

\begin{figure*}[htbp]
\centering
\includegraphics[width=\textwidth]{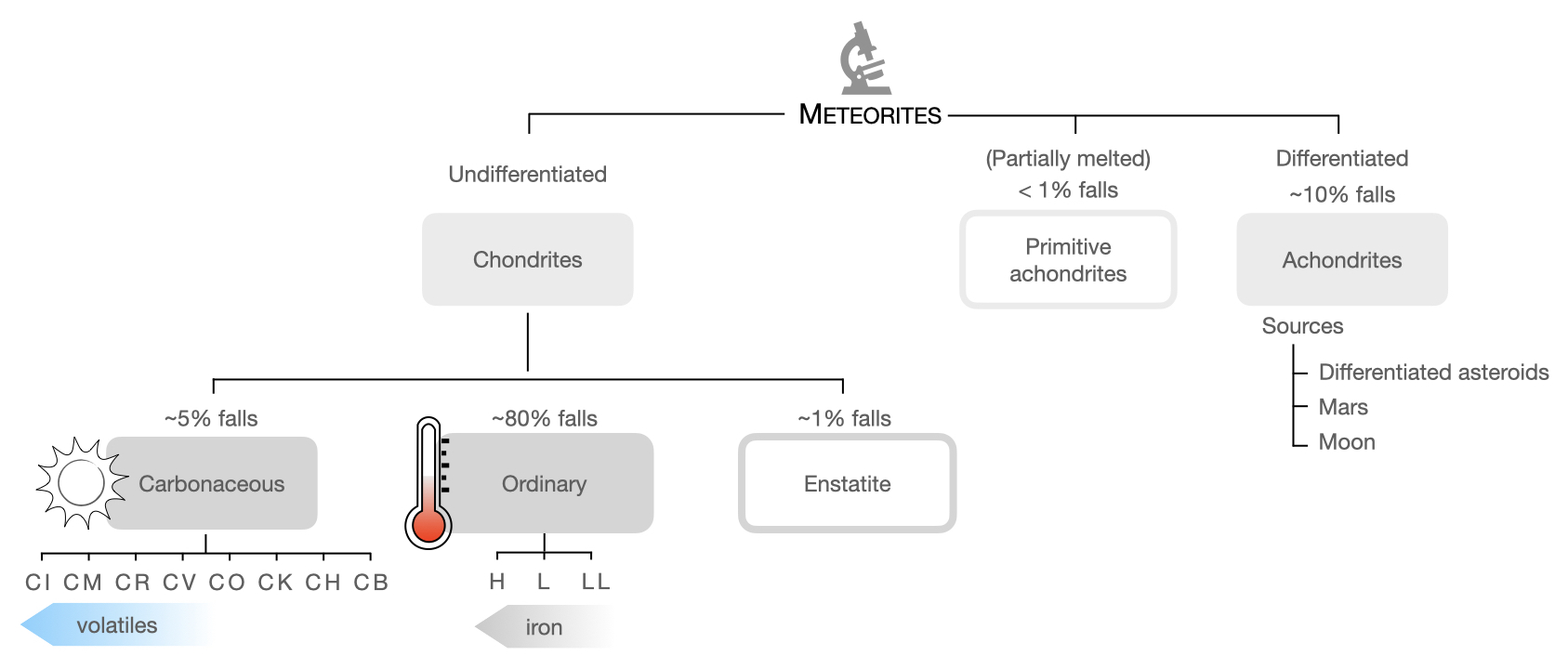}\hspace{5pt}
\caption{Major classes of meteorites relevant to sample return missions, adapted from \citet{Weisberg2006}} 
 \label{fig:meteorite_classification}
\end{figure*}

Meteorites are natural samples of celestial rocky material delivered to Earth. Observations of their trajectory during atmospheric entry suggest that the majority of meteorites originate from asteroids, which are most densely concentrated in the asteroid belt between the orbits of Mars and Jupiter. Much of the present‑day asteroid population consists of fragments of original parent bodies that were catastrophically disrupted and later modified by subsequent collision events. Their orbits can be perturbed towards the Earth by thermal effects and the gravitational influence of the planets. Larger bodies may become Near Earth Objects (NEOs) which can present a danger if their orbits become Earth-crossing. However, small fragments typically fall harmlessly through our atmosphere to land on the Earth's surface as meteorites. 

Meteorites can be divided into classes based on geological processes and composition (Figure~\ref{fig:meteorite_classification}). One of the fundamental distinctions is between `chondrite' and `achondrite' meteorites. Achondrites are fragments of a celestial body that has undergone some degree of melting on a planetary scale, resulting in the process of differentiation whereby denser materials such as iron metal separate from the lighter silicates. Differentiation is the first step towards a geologically active planet, changing the internal structure of a celestial body from a homogenous mix of elements to the core, mantle, and crust layers of a terrestrial world. Differentiation occurs in celestial bodies larger than approximately 30\,km, so meteorites originating from Mars, the Moon, or larger asteroids are achondrites.  

By contrast, chondrites consist of material that has never experienced the large-scale planetary melting process of the achondrites and remain undifferentiated. The bulk composition of chondrites has a relative abundance of non-volatile chemical elements that closely matches that of the solar photosphere. This similarity indicates that chondrites formed in the early Solar System from the same nebula material that was  forming the young Sun. Chondrites are therefore considered `primitive' meteorites, representative of the ancient material that accreted to build the Earth and other planets. A key consequence is that radiometric dating of chondrites provides an age for the first rocky material in the Solar System. The oldest material within chondrites are submillimeter- to centimeter-sized inclusions known as chondrules, Ca-Al-rich inclusions (CAIs), and amoeboid olivine aggregates (AOAs), which are thought to have formed at high temperatures in the inner regions of the solar nebula. The age of CAIs is considered the $t = 0$ for the Solar System, and is recorded at approximately 4.6 billion years.

Although all chondrites are of undifferentiated primitive material, their precise composition depends on the condensible elements in the region of the solar nebula where they formed, and the accretion and subsequent evolution of their asteroid source body. These compositional variations divide chondrites into types (and sub-types). The most common meteorite discoveries are `ordinary chondrites', and account for approximately 80\% of all observed meteorite falls. Ordinary chondrites typically have a low volatile content and many have experienced high temperatures leading to thermal metamorphism (changes to the mineral composition or solid state texture due to heat). The class is subdivided into H, L, and LL chondrites, denoting H(igh) iron, L(ow) iron, and LL(ow) iron and metal abundances, and associated increasing iron oxidation state, from predominantly Fe$^0$ metal in H chondrites, to mostly oxidised Fe$^{2+}$ within silicates in LL chondrites.

Amongst the most uncommon meteorite finds are `carbonaceous chondrites'. Carbonaceous chondrites differ markedly from ordinary chondrites, with the closer compositional match to the solar photosphere and often retain substantial volatile abundances, both indicative of a lower degree of thermal processing since formation. The volatile inventory can include significant amounts of water chemically bound within the structure of the minerals of the rock, especially the CI, CM, and CR chondrite sub-classes (highlighted in Figure~\ref{fig:meteorite_classification}). Minerals with water bound within their structure are known as hydrated minerals, for which a common example on Earth is clay. Carbonaceous chondrites may also contain organic matter. This inclusion of molecules needed for life on Earth has lent particular interest to carbonaceous chondrites as providing possible evidence for how our planet became habitable. 

The rarest subtype of carbonaceous chondrite is the CI chondrites, also referred to as Ivuna-like carbonaceous chondrites. Over 77,000 meteorites are registered in the Meteoritical Bulletin database but only ten are examples of CI chondrites, including the Ivuna meteorite that fell in Tanzania and lends its name to the type. CI chondrites have the highest abundance of volatiles of all meteorite types, and a composition that is most similar to the solar photosphere. Until the return of the Hayabusa2 asteroid sample return mission, CI chondrites were considered the most chemically pristine reference material for the original composition of the Solar System. 

Despite providing valuable insights into the formation of the Solar System, meteorite studies face three important limitations. The first is that a meteorite is a sample of an unknown asteroid. Without being able to observe the source celestial body, it is not possible to know if the meteorite represents the asteroid bulk properties, or is a sample of one of several lithologies (e.g. rock type). Although chondrites are undifferentiated, the scale of compositional homogeneity will vary depending on the history of the asteroid. As we will see, interpretation of the returned samples from the asteroid sample return missions is closely linked to the remote-sensing observations made by the spacecraft. Meteorites also do not provide information about the orbit of their source asteroid, removing information that could help trace the path of this material through the Solar System. Meteorites are therefore samples without geological context, and extrapolating their properties to broader questions about the transportation and evolution of material in the Solar System is extremely challenging.

The second problem is that the passage through the Earth's atmosphere is highly destructive. Intense aerodynamic heating created during the fall of a solid object passing at high speed through the air will vaporise material. This risks a bias in the meteorite population that reaches the ground, as more robust materials will be favoured. Although volatile material may be preserved within the interior of surviving meteorites, fragile samples are unlikely to survive atmospheric entry. Evidence for this selection effect will be seen in the match between two of the returned asteroid samples and the rarest of meteorite discoveries.

The third issue is contamination from the Earth's environment. Even meteorites rapidly recovered after observed falls show changes due to terrestrial alteration. For example, the Winchcombe meteorite landed on a driveway in the UK on 28 February 2011, and fragments were recovered within hours to days and stored in sealed bags. Yet, table salt (halite) was still found within the sample \citep{Jenkins2024}. The incorporation of terrestrial molecules and any subsequent chemical reactions compromise the compositional record preserved in the meteorite. These limitations mean that meteorites alone cannot fully constrain the conditions in the early Solar System, nor provide a complete picture of the Earth's origins.

\subsection{Ground-based observations}
\label{sec:groundbased}

\begin{figure*}[!t]
\centering
\includegraphics[width=\textwidth]{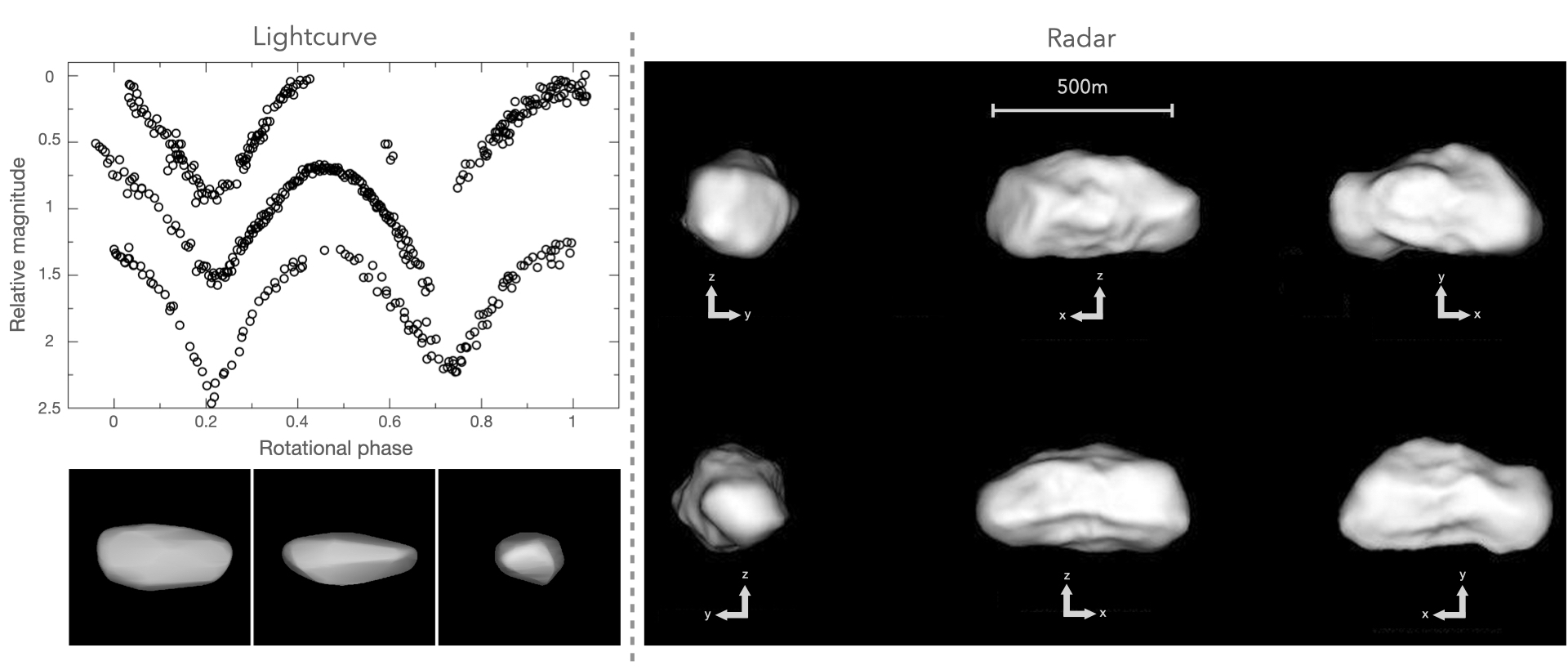}\hspace{5pt}
\caption{Ground-based observations of asteroid Itokawa before the arrival of the Hayabusa spacecraft. The left-hand images show three composite lightcurves of Itokawa above the corresponding shape model (reproduced with permission from \citet{Kaasalainen2003} \raisebox{0.3ex}{\scalebox{0.8}{\copyright}}\,2003 ESO),  while the right-hand image shows the shape model derived from radar observations (reproduced with permission from \citet{Ostro2005} \raisebox{0.3ex}{\scalebox{0.8}{\copyright}}\,2010 John Wiley \& Sons).} 
 \label{fig:itokawa_lightcurve}
\end{figure*}

\begin{figure*}[!t]
\centering
\includegraphics[width=\textwidth]{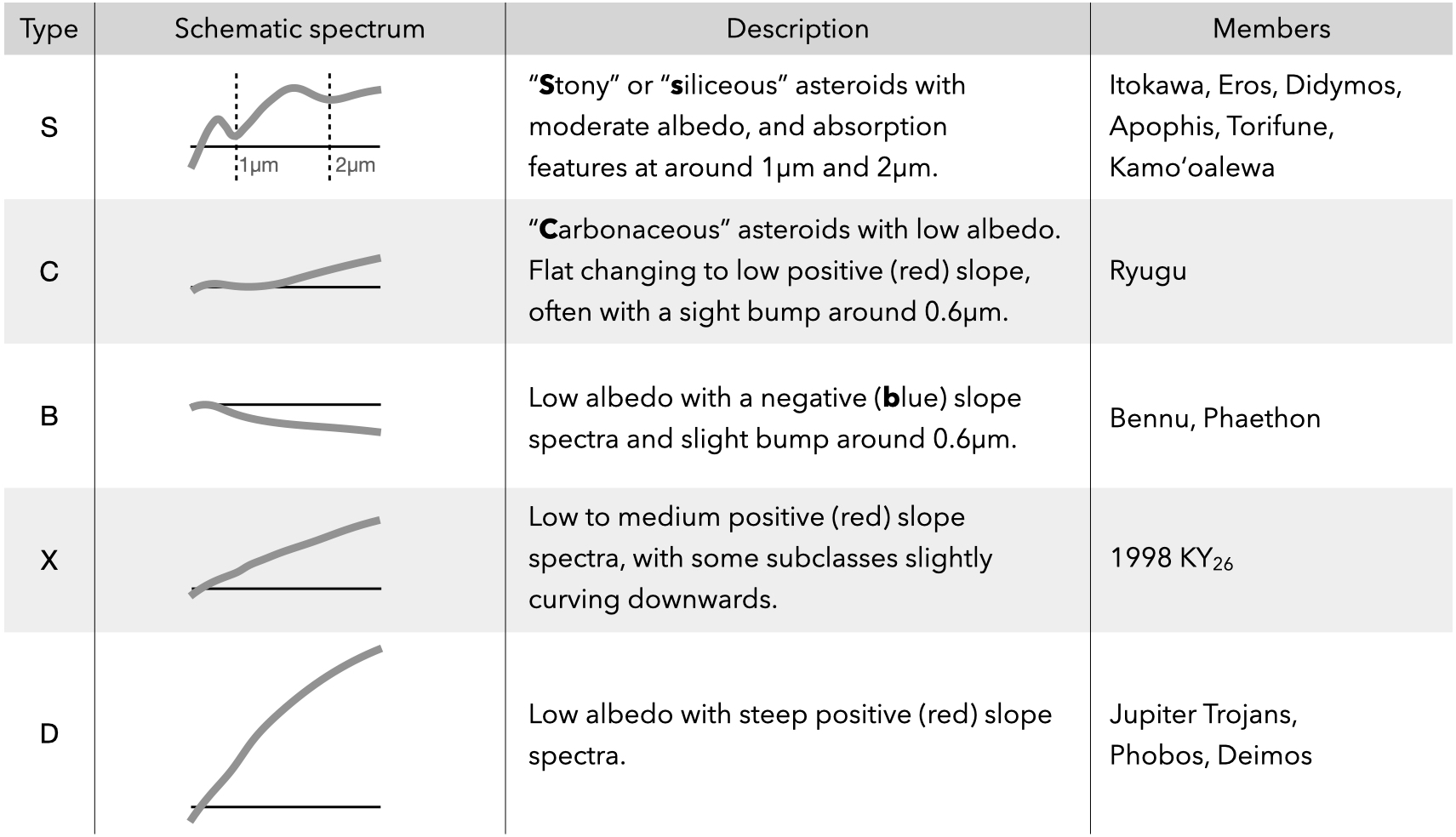}
\caption{Asteroid classification based on reflectance spectra across visible to near-infrared wavelengths (0.45 to 2.45\micron). These five asteroid types are major examples in the Bus-DeMeo classification system, consisting of 24 asteroid types and subtypes \citep{DeMeo2009}. The two Martian moons, Phobos and Deimos, are listed here due to the similarity of their spectra with D-type asteroids. However, their origin is currently unknown. The schematic spectra are adapted from \citet{DeMeo2009}, and the bold lettering indicates the origin of the name.} 
\label{fig:asteroid_classification}
\end{figure*}

The word `asteroid' originates from the Greek meaning `star-like', a definition that indicates the main challenge of ground-based observations. Even when viewed through powerful telescopes with high magnification, asteroids appear only as point sources of light. For example, if asteroid Ryugu (with a diameter of approximately 1\,km) were to approach the Earth as close as the Moon, the apparent size when viewed from Earth would be just 0.5\,arcsec, where 1\,arcsec is 1/3600 degrees. In contrast, the Moon with a diameter of 3,500\,km has an apparent size of 0.5 degrees. Even the 8.2\,m Subaru Telescope in Hawai'i would still be unable to resolve asteroid Ryugu at that proximity.

Ground-based observations therefore primarily observe variations in the brightness over time (photometry) and the intensity across different wavelengths (spectrometry) of the reflected light from an asteroid. Changes in the brightness occur as the asteroid rotates and different surface features reflect light towards the Earth. These variations are captured by lightcurves (graphs of brightness magnitude plotted against time) that can be analysed to determine the asteroid rotation rate, orientation of the rotation axis, and to estimate the relative dimensions. Lightcurves for asteroid Itokawa from photometric observations conducted during 2001 and the resulting shape model are shown in Figure~\ref{fig:itokawa_lightcurve}.

If an asteroid passes sufficiently close to the Earth to reflect radio waves from the surface, then radar observations can be used to measure these properties at much higher resolution than optical methods. This technique requires a powerful radio transmitter and a large receiving antenna to detect the attenuated return signal. For over 50 years, the 350\,m Arecibo Telescope was the primary facility for asteroid radar observations until major structural damage in 2020 ended operations. The main work horse for this task is currently the NASA Goldstone Solar System Radar. Radar data can generate significantly more accurate shape models of asteroids compared to photometry, but neither technique can directly visualise the surface environment. The shape model from radar observations of asteroid Itokawa is shown alongside that from the lightcurves in Figure~\ref{fig:itokawa_lightcurve}. Of the three asteroids visited by sample return missions, asteroids Itokawa and Bennu both were observed with radar. Asteroid Ryugu did not approach the Earth sufficiently closely for radar observations to be conducted.

Plotting the intensity of light reflected across different wavelengths gives the asteroid's spectrum. A spectrum is described as `red' when more light is reflected at longer wavelengths and the plotted spectrum has a positive slope, and `blue' when the reflection is stronger at shorter wavelengths to give a negative slope. Since different minerals and molecules absorb and reflect specific wavelengths of light, the shape of spectrum should relate to the surface composition of the asteroid. The spectral profile is therefore used as a basis for asteroid classification. The Bus-DeMeo taxonomic classification system divides the observed asteroid population into 24 types and subtypes based on the shape of their spectral profile across visible and near-infrared wavelengths \citep{DeMeo2009}. Figure~\ref{fig:asteroid_classification} shows a schematic of the spectrum of five major types of asteroid, including the three types visited by asteroid sample return missions and those for up-coming flyby, rendezvous, and sample return missions. 

There is a general trend between asteroid type and the distance of the asteroid orbit from the Sun. S-type asteroids are most commonly found in the inner asteroid belt, between approximately 2.1 and 2.5\,au. C-type asteroids are significantly more common in the outer belt, between about 2.8 and 3.3\,au, and are the most numerous asteroid class overall. B-type asteroids (often grouped within a broader C-type complex), and X-type asteroids follow a similar distribution to C-type asteroids, with higher abundances in the outer belt. Meanwhile, D-type asteroids predominantly exist in the outer Solar System, with the majority of the Trojan asteroids that share the orbit of Jupiter being classified at D-type. The two moons of Mars, Phobos and Deimos, share spectral similarities with D-type asteroids, although the origin of the pair is currently undetermined. 

The different orbital distances of the asteroid classes supports a connection between spectral taxonomy and composition, as the temperature gradient enables the condensation of different materials at varying solar distances. However, the actual properties of the asteroids within each spectral type is very uncertain. This is due to the difficulty in matching remote spectroscopic observations of asteroids with microscopic analyses of their meteorite fragments. Studies based on the asteroid albedo, orbital location, and comparison with meteorite spectra suggested that C-type, B-type, and D-type asteroids are likely to have primitive compositions that are rich in organics and hydrated minerals, whereas S-type asteroids appear to be more volatile-poor and may have experienced significant heating. A definitive compositional match can only be made by delivering samples from an asteroid back to Earth for laboratory analysis. This has been achieved for three asteroids, but--as we will see--these sample return missions have also revealed additional complications in interpreting the remote data due to surface alterations through processes such as space weathering. 

\subsection{Space-based remote-sensing missions}

Observations from the spacecraft in the vicinity of the asteroid have a crucial role in sample return missions by providing essential geological context for interpreting the laboratory analyses of the returned materials. Even when a physical sample cannot be returned, high spatial resolution data collected by the spacecraft offers a wealth of information unavailable from the ground. 

Remote-sensing missions can be either a flyby of the asteroid, or rendezvous. A rendezvous mission allows extended observations at close proximity, but at the cost of high energy requirements (represented by the necessary change in velocity, $\Delta V$) needed to manoeuvrer the spacecraft into an orbit that matches that of the target asteroid. For targets that require a $\Delta V$ too high to fit the mission specifications, an asteroid flyby offers a single chance opportunity to collect data while passing the asteroid at high speed. 

The first mission to rendezvous with an asteroid was the NASA Near Earth Asteroid Rendezvous (NEAR) Shoemaker spacecraft, which entered into orbit around the large (16.8\,km diameter) S-type asteroid 433 Eros in 2000. Rendezvous missions like NEAR Shoemaker enable detailed mapping of an asteroid's shape and surface morphology, take measurements of the surface composition, and can monitor rotation and temporal variations. The effect of the gravitational pull of the asteroid on the spacecraft can also be used to calculate the asteroid density and estimate internal structure. NEAR Shoemaker discovered that Eros is a consolidated body with a density of 2.67\,\gcm, with a surface coated with a thick layer of regolith formed by micrometeorite impacts. Detailed spectral data revealed a silicate mineralogy that was consistent with ordinary chondrite meteorites, although the measurements were insufficient to establish a link with a specific meteorite type. NEAR Shoemaker also detected no intrinsic magnetism \citep{Cheng2002}. 

Prior to arriving at Eros, NEAR Shoemaker performed a flyby of C-type asteroid 253 Mathilde in 1997. This encounter returned the first resolved images of Mathilde, although only one face could be imaged due to the brief duration of the flyby. Based on the spacecraft tracking data and images, the density of Mathilde was estimated to be low and suggestive of a highly porous and possibly rubble-pile structure. The challengingly brief data collection opportunity presented by a flyby is offset by lower launch costs and can be integrated into mission plans with broader objectives. For example, the NASA LUCY mission and up-coming JAXA DESTINY+ mission will perform multiply flybys to study several asteroids. As demonstrated by NEAR, flybys can often be incorporated into existing mission designs at minimal additional cost. The first asteroid flybys were of S-type asteroids 951 Gaspra and 243 Ida by the NASA Galileo mission en route to Jupiter, and led to the discovery of the first asteroid moon, Dactyl. The extended missions of JAXA Hayabusa2 and NASA OSIRIS-REx after their sample return both include asteroid flybys.

While remote analysis yields valuable data, the range and precision cannot match that of  Earth-based laboratories that are not constrained by spacecraft operations. Returned samples also offer long-term availability, allowing further analysis as understanding evolves. A percentage of the returned sample is typically preserved for longer-term future research. For example, the final vacuum-sealed lunar sample from the Apollo missions was only opened in 2022, nearly fifty years after Apollo 17 returned that material to Earth. Similarly, each of the asteroid sample return missions have set aside part of the collected material for future generations of scientists.

\vspace{0.2cm}
Asteroid sample return missions have the potential to integrate data from meteorite studies, ground-based observations, and spacecraft remote-sensing to provide a ground truth for asteroid properties. The data offers insights into the formation and evolutionary history of asteroids, which in turn provides evidence for the earliest processes that shaped our planet. 

\section{The Hayabusa mission: sample delivery 2010}

The first asteroid sample return mission was launched by JAXA from the Uchinoura Space Center in Japan on 9 May 2003. The mission successfully returned to Earth with a sample from the S-type asteroid Itokawa on 13 June 2010.

Originally designated MUSES-C, the spacecraft was renamed `Hayabusa' (`falcon' in Japanese) after reaching orbit. As the first attempt to deliver material from an asteroid, Hayabusa was primarily a technology demonstration mission for small body exploration and sample acquisition. The mission tested several key technologies that included ion engine propulsion for the interplanetary travel, autonomous navigation using optical imaging, autonomous descent and sample collection in a microgravity environment, and the direct return of a sample capsule from interplanetary space.

\begin{figure*}[!t]
\centering
\includegraphics[width=\textwidth]{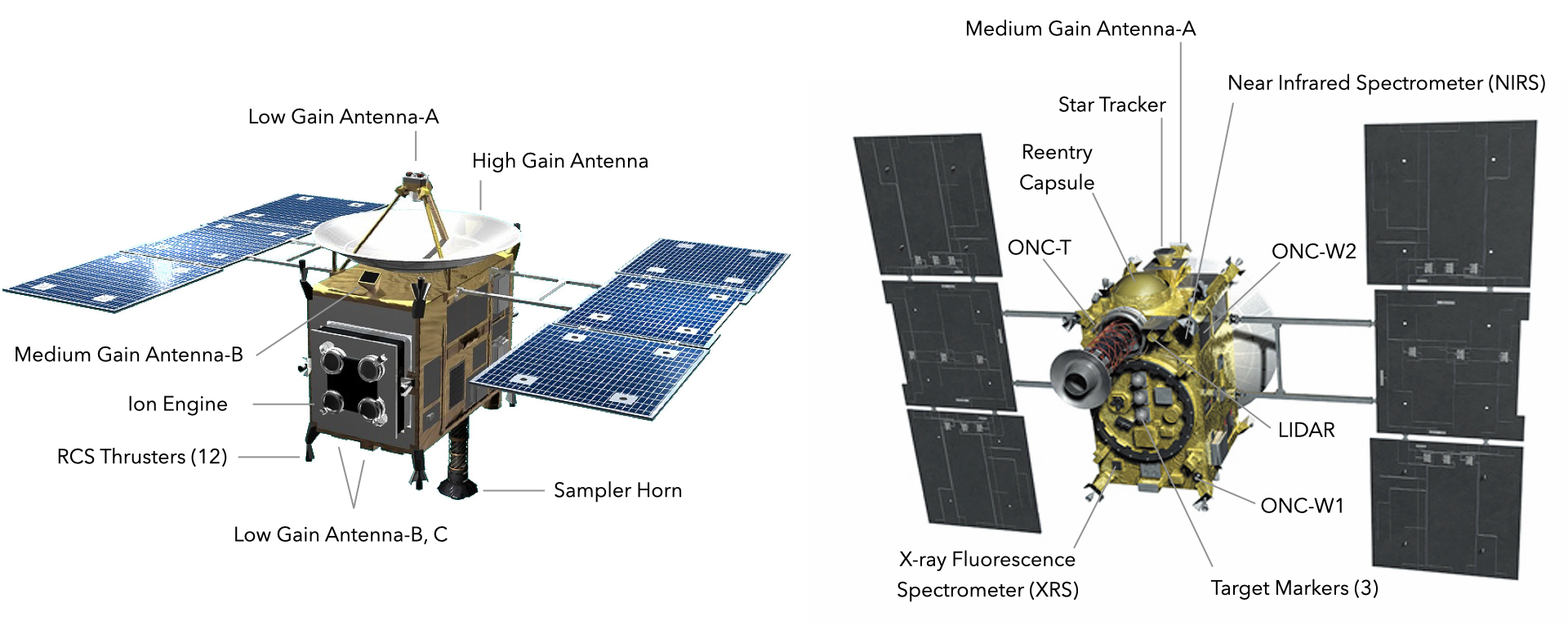}
\caption{Diagram of the Hayabusa spacecraft (credit: JAXA).} 
\label{fig:haya_spacecraft}
\end{figure*}

In addition to the technology specifically for the sample return, the Hayabusa spacecraft carried a suite of remote sensing instruments for characterising asteroid Itokawa during proximity operations (Figure~\ref{fig:haya_spacecraft}). These included two wide-view cameras for optical navigation (ONC-W1, -W2), a laser altimeter (LIDAR) for precise range measurements and surface topography mapping, and the telescopic Asteroid Multi-band Imaging Camera (AMICA or ONC-T) equipped with both a wide bandpass filter for imaging and navigation and seven narrowband filters for multispectral imaging across far-ultraviolet to near-infrared wavelength range (0.38 – 1.01\micron), a near-infrared (0.85 - 2.1\micron) spectrometer (NIRS), and an X-ray fluorescence spectrometer (XRS) all used for compositional analysis of the asteroid surface.

\subsection{The mission}

The flight of Hayabusa would make space history, not only for returning the first asteroid sample to Earth, but also for overcoming extraordinary challenges that would later inspire three movies and a documentary in Japan. Just months after launch, the spacecraft was struck by one of the largest solar flares in recorded history. The resulting burst of energetic particles from the Sun degraded the solar cells and damaged critical systems including reaction wheels, ion engines, the chemical propulsion system, and the sampling mechanism. Creative solutions were required throughout the mission to compensate for the failing hardware, enabling Hayabusa to collect the asteroid sample and return home (an excellent mission account is also given in \citet{Yoshikawa2015}). 

The Hayabusa spacecraft was designed to collect asteroid material through a 1\,m-long sampler horn, sized to prevent the spacecraft's solar array panels making contact with the surface during touchdown (Figure~\ref{fig:haya_spacecraft}). During the descent to the asteroid surface, a target marker was released to provide a reflective reference point. The apparent motion of the target marker relative to the asteroid surface was tracked by the onboard cameras and used to cancel the spacecraft's lateral velocity to ensure a gentle touchdown. In the planned sampling sequence, a projectile would fire down the sampler horn upon contact with the surface and stir the surface material or break up the rocks. Ejecta released in the impact would then rise through the sampler horn in the microgravity environment and into sample container. However, the damage to the spacecraft prevented the touchdown operations completing as expected. 

Hayabusa attempted two sample collections on the 20 and 25 November 2005 JST. During the first touchdown, a fault with the obstacle detection sensor caused the spacecraft to become stranded on the asteroid surface for almost thirty minutes. The second touchdown progressed more smoothly, but the projectile to lift surface material was not fired during either operation. 

Hayabusa departed from asteroid Itokawa in April 2007 after collecting extensive remote sensing data but without the certainty that any sample had been successfully acquired. During the return journey, a gas eruption from the damaged chemical thrusters sent the spacecraft tumbling and communication with the ground was lost for seven weeks. 

Three years later than originally scheduled, Hayabusa reached Earth with no fully functional ion engine. An ion engine consists of a source of xenon ions that are accelerated into a beam, and a neutraliser that emits electrons to cancel the ion charge before the spacecraft becomes electrically charged. With none of the four engines possessing both a working ion source and neutraliser, the final journey was achieved through an ingenious cross-connection of the components from two engines. 

The sample container onboard Hayabusa was stored within a 20\,kg reentry capsule, which successfully separated from the damaged spacecraft. However, it was no longer possible to adjust the trajectory to prevent the spacecraft from following the capsule into the Earth's atmosphere. The team sent a final command to rotate Hayabusa so that the camera could see Earth for a final time before the spacecraft plunged into the atmosphere three hours after the reentry capsule and disintegrated.

The separated reentry capsule was equipped with a thermal shield to protect the sample from intense aerodynamic heating during atmospheric entry. The shield was jettisoned to allow deployment of a parachute for the final descent to the Woomera Prohibited Area in South Australia. The landing location of the reentry capsule was identified by the ground retrieval operation team using a beacon transmitter embedded in the capsule and was discovered within 30 minutes of atmospheric entry. The capsule was then placed in a container filled with pure nitrogen to prevent terrestrial contamination during the transportation back to Japan. 

Multiple precautions were taken to minimise the possibility of contaminating the sample with terrestrial materials (see \citet{Yada2014} for a detailed description). These measures included constructing the sample container from a limited set of chemically inert materials, ultrasonic cleaning in isopropyl alcohol (IPA) to remove organic residues, installation onboard the spacecraft in a class 10,000 clean room ($\le10,000$\, airborne particles/ft$^3$ larger than 0.5\micron), and adding a contamination coupon to monitor any potential pollutants. Once back on Earth, the sample container was transported to cleanrooms within the JAXA Extraterrestrial Sample Curation Center with classification between class 100 and class 1000, and featured stainless steel 304 clean chambers that have been baked to reduce any contaminant gases. The sample container was opened in a clean chamber under vacuum conditions, and then moved to a second chamber for analysis in an environment of pure nitrogen.

It was uncertain whether any sample had been collected during the two attempts on Itokawa. But when the sample container was opened in Japan, thousands of small particles were found clinging to the inside surface of the two catcher rooms that corresponded to the two touchdowns. A Teflon spatula was used to sweep particles from about 10\% of the container, but ultimately the most successful retrieval method was to physically tap the container. An exact mass for the total sample collected by Hayabusa is not recorded, but an approximate value is 100\micron, comprising of more than 1500 particles smaller than about 180\,\micron. Analysis of the oxygen isotopic composition of these grains revealed a clear distinction from terrestrial material, confirming that this was the first sample ever collected from an asteroid \citep{Yurimoto2011}. Laboratory studies continue worldwide, demonstrating the value of even a tiny returned sample.

\subsection{Asteroid Itokawa}

\subsubsection{First impressions: the `otter' asteroid}

The target asteroid for the Hayabusa mission was the near-Earth asteroid (25143) Itokawa. The asteroid was discovered in September 1998 by the Lincoln Near Earth Asteroid Research (LINEAR) program and given the provisional designation 1998 SF$_{36}$. In 2003, the International Astronomical Union officially approved the name `Itokawa', following a request by LINEAR on behalf of the Hayabusa mission team. The asteroid is named for aerospace engineer Hideo Itokawa (1912-1999), who is regarded as the father of Japanese rocketry. 

The orbital path of Itokawa extends from just inside the Earth's orbit at 0.95\,au to a short distance past the orbit to Mars at 1.69\,au (where 1\,au is the average Earth - Sun distance). As an engineering mission, the  selection of asteroid Itokawa was driven primarily by accessibility based on the planned launch date and available propulsion capabilities. 

\begin{figure*}[!th]
\centering
\includegraphics[width=\textwidth]{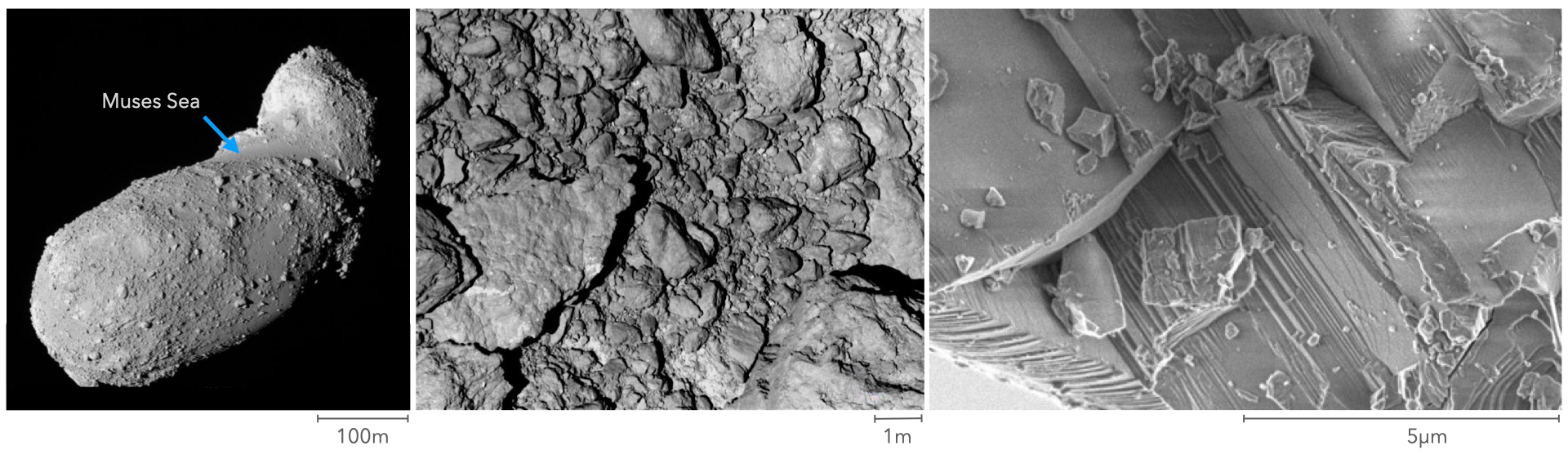}
\caption{Images of asteroid Itokawa. Left and centre images were captured by the spacecraft while in the vicinity of the asteroid, while the right-hand image shows microparticles from the returned sample photographed with an electron microscope. Sample return missions allow high resolution images from global down to microscopic (credit: JAXA).} 
\label{fig:itokawa}
\end{figure*}

Between September and early December 2005, Hayabusa conducted close proximity observations of Itokawa. The prior ground-based lightcurve and radar studies had indicated that Itokawa had an ellipsoid shape (see Figure~\ref{fig:itokawa_lightcurve}), which was confirmed when the spacecraft returned images of an asteroid that resembled a floating sea otter, comprising of a smaller, rounded `head' and larger `body' (Figure~\ref{fig:itokawa}). The longest axes on the asteroid measures 535\,m, with orthogonal dimensions of 294\,m and 209\,m for the shorter axes. Approximately 80\,\% of the asteroid surface is covered by rough terrain that includes numerous boulders, while two smoother areas of fragmented cm- to mm-sized grains extend around the `neck' region between the head and body. Despite these morphological contrasts in structure, remote-sensing observations with the NIRS near-infrared spectrometer revealed no significant variations in the mineralogical composition across the surface \citep{Abe2006}. The two sampling sites were selected about 100\,m apart in the smooth neck region that was later dubbed `Muses Sea' after the original designation of the mission. 
 
\subsubsection{The link between S-type asteroids and LL-chondrites}

Ground-based observations of the asteroid spectrum had classified Itokawa as an S-type asteroid, which is a common taxonomic group in the inner asteroid belt (see section~\ref{sec:groundbased}). The average spectrum from the NIRS spectrometer onboard Hayabusa confirmed this classification, detecting the broad absorption band at 1\,\micron that is characteristic of the S-type class (see Figure~\ref{fig:asteroid_classification}) \citep{Abe2006}. 

\begin{table}[t]
\caption{Abundance (by weight percentage, wt.\%) of the main minerals in the
Itokawa returned sample \citep{Nakamura2011, Yoshikawa2015}.
Bold indicates the dominant chemical composition.}
\label{table:itokawa_minerals}
\vspace{5pt}
\centering
\begin{tabular*}{\columnwidth}{@{\extracolsep{\fill}} l l r}
\hline
Mineral & Formula & wt.\% \\
\hline
Olivine & (\textbf{Mg},Fe)$_2$SiO$_4$ & 64.9 \\
Low-Ca Pyroxene & (\textbf{Mg},Fe,Ca)SiO$_3$ & 18.6 \\
High-Ca Pyroxene & (\textbf{Ca},Mg,Fe)SiO$_3$ & 2.8 \\
Plagioclase & (\textbf{Na},Ca)Al(Al,Si)Si$_2$O$_8$ & 11.2 \\
Troilite & FeS & 2.1 \\[1ex]
\hline
\end{tabular*}
\end{table}

Analysis of the delivered samples revealed that the collected asteroid grains have an average density of 3.4\,\gcm. Material collected from the two sampling sites exhibit no significant difference in petrology, and have a largely homogenous chemical composition. This narrow range in composition is indicative of a period of intense thermal metamorphism in Itokawa's past, during which high temperatures would have allowed the minerals to recrystallise into a state of close chemical equilibrium. The composition consists primarily of the silicate minerals olivine, low- and high-Ca pyroxenes, and plagioclase, and the iron-sulphide troilite (see Table~\ref{table:itokawa_minerals}). These characteristics--composition, density and evidence of thermal metamorphism--closely match that of the LL ordinary chondrite meteorites \citep{Nakamura2011}. While the link between S-type asteroids and LL chondrites had been suspected, these samples provided the first definitive confirmation of the link between the spectral class and meteorite group. 

\subsubsection{Space weathering}
\label{sec:haya_spaceweathering}

The Itokawa sample had established the relationship between LL chondrites and S-type asteroids but questions still remained. In particular, S-type asteroids exhibit a redder spectrum than ordinary chondrites, which had previously cast doubt on their common origin. Hints as to the source of this discrepancy were found in the spatially resolved observations of Itokawa by Hayabusa. Although the average reflectance spectrum affirmed the S-class classification of the asteroid, the colour and albedo are not homogenous. Steep slopes where erosion can expose fresh material appear bluer and brighter, while less disturbed regions are redder and darker. These variations were suspected to be due to space weathering: a surface-altering process that occurs on airless bodies primarily triggered by the solar wind, cosmic rays, and micrometeorite impacts. Space weathering was first recognised in lunar samples, where the upper layer is darker and redder because of deposits of extremely small iron particles. On the Moon, these deposits are primarily produced when micrometeorite impacts vaporise material, which then condenses into a layer rich in nanophase metallic iron. If the same reddening process was occurring on asteroids, then this could explain the difference between S-type asteroids and their ordinary chondrite counterparts.

The sample returned from Itokawa did exhibit similar layers of nanophase iron-rich particles. However, the weak gravity of the asteroid hinders the retention of vapour, and the thin, mobile surface regolith does not allow impact damage to accumulate. Reddening on Itokawa is therefore driven the solar wind rather that micrometeorites. High-energy ions in the solar wind implant into the mineral surfaces, reducing Fe$^{2+}$ ions to metallic Fe$^0$, and disrupting the crystal lattice to concentrate Fe$^0$ through radiation-induced segregation \citep{Noguchi2014}. As on the Moon, the result is a reddening of surface material over time due to the build-up a nanophase iron-rich deposit, but driven by the solar wind rather than micrometeorite impacts.

This was the first direct evidence of space weathering on an asteroid and had profound implications for our understanding of how asteroid surfaces, like the Moon, may evolve over time. The weathering explains observed differences in spectra between S-type asteroids and ordinary chondrite meteorites, and also demonstrates how space weathering can operate on small bodies. 

\begin{figure*}[!th]
\centering
\includegraphics[width=\textwidth]{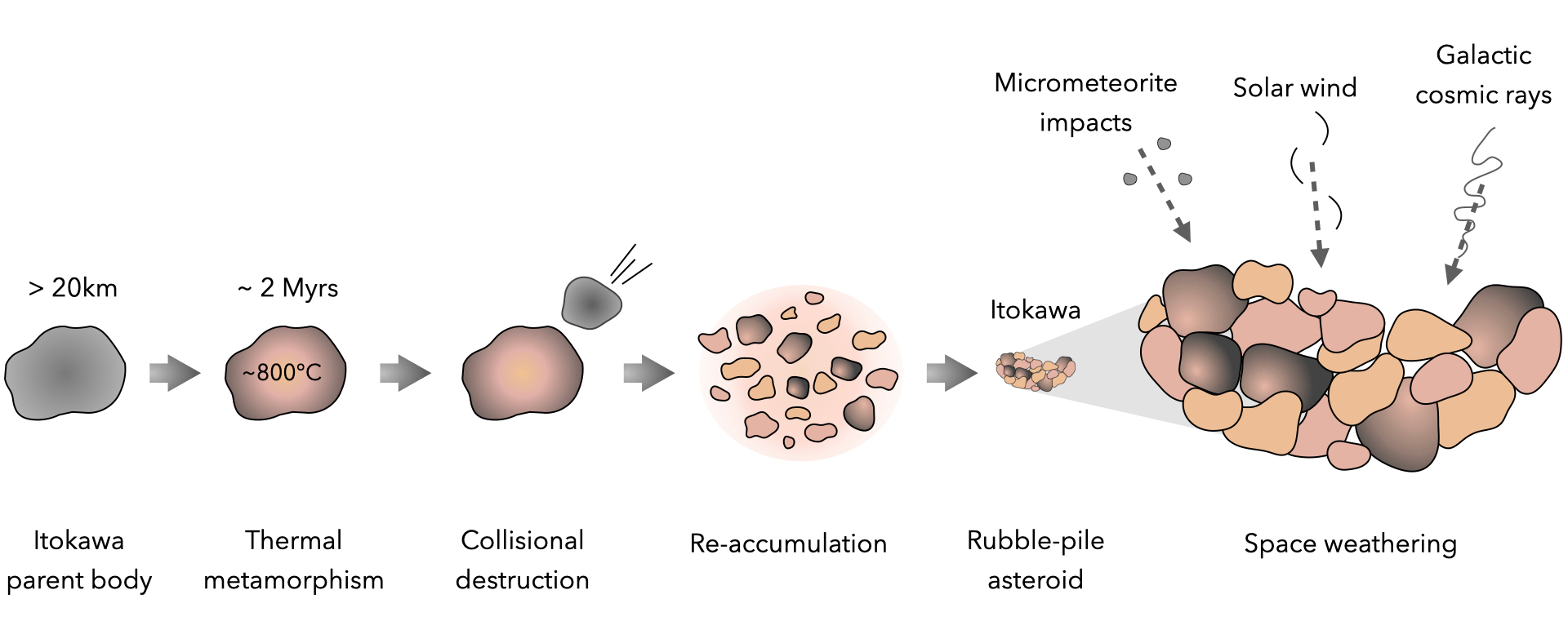}
\caption{The history of asteroid Itokawa.} 
\label{fig:itokawa_history}
\end{figure*}

\subsubsection{A rubble pile asteroid}

The mass of asteroid Itokawa was determined by analysing the trajectory of the Hayabusa spacecraft during a series of close approaches to within 100\,m of the asteroid surface \citep{Fujiwara2006}. Small deviations in the motion of the spacecraft caused by the gravitational pull from Itokawa indicated that the asteroid has a mass of $3.51\times 10^{10}$\,kg. The volume of the asteroid was derived from the three-dimensional shape model constructed from images captured by the spacecraft cameras, yielding an estimate of $1.84\times 10^7$\,m$^3$. Combining these values gives a bulk density for Itokawa of 1.9\,\gcm. 

This asteroid bulk density is considerably lower than the grain density of the sample, which was measured in the laboratory as 3.4\,\gcm. The macroporosity of Itokawa is therefore around 40\,\%, suggesting the asteroid is not a monolithic body, but a loose rubble pile with an interior that includes substantial voids. This interpretation is supported by the absence of any large-scale structural features, such as visible ridges or long linear formations that extend across the length of the asteroid, suggesting that Itokawa is not a single consolidated body but consists of fragments that are not greater than 50\,m in size \citep{Fujiwara2006}. The existence of rubble pile asteroids had previously been hypothesized, but Itokawa was the first direct observational confirmation of their existence. 

\subsubsection{The history of an NEO}

Evidence for how the rubble pile structure was created lie in the asteroid shape and composition. The high-resolution images returned by the spacecraft suggest that Itokawa was once composed of two separate small bodies that merged to form the current configuration \citep{Saito2006}. The head and body components of the sea otter shape have different orientations and are connected by a narrow neck with a steep gradient. While these two sections could have been independent asteroids, this scenario is unlikely because the probability of an encounter at sufficiently low relative velocities to avoid collisional breakup is very small. Furthermore, the remote sensing data revealed no difference in the mineralogy between the two regions, indicating a common origin. It is therefore more probable that Itokawa was once part of a larger parent body that underwent catastrophic disruption. The flying fragments from this event re-accumulated into rubble pile bodies, two of which began to orbit one another as a binary system before coming into contact \citep{Fujiwara2006}. Ground-based observations have identified numerous elongated and binary asteroids, suggesting that the formation of Itokawa may represent a common evolutionary pathway for many small bodies (Figure~\ref{fig:itokawa_history}). 

Clues to the evolution of the parent body that fragmented to create Itokawa are preserved in the minerals of the returned sample grains. The particular crystalline structure of the plagioclase and the homogeneous compositions of the pyroxene indicate that these minerals formed at peak temperatures of approximately 800\degC. Additional constraints come from the olivine-spinel geothermometer, which estimates temperature based on the (temperature dependent) distribution of iron and magnesium in sample grains containing both olivine and chromite (a spinel group mineral). This method yielded temperatures near 600\,\degC, implying that the parent body cooled slowly to this level, allowing sufficient time for Fe-Mg exchange between olivine and chromite to reach chemical equilibrium at this temperature \citep{Nakamura2011}. The evidence for high temperatures indicates that Itokawa is best classified as an LL5-6 ordinary chondrite, among the most thermally metamorphised petrologic types. 

The prolonged high temperatures recorded by the Itokawa minerals are most plausibly attributed to the heat generated by the decay of short-lived radioisotopes such as $^{26}$Al. Meteoritic evidence indicates that $^{26}$Al was abundant in the early Solar System, serving as the primary source of internal heat for the rocky planetesimals that first formed around the Sun. Larger celestial bodies and those that formed earlier during greatest abundances of $^{26}$Al would have incorporated larger quantities of the radioisotope and reached higher internal temperatures. Based on the measured peak temperature and evidence for slow cooling, Itokawa likely originated from a parent body larger than 20\,km in radius that accreted approximately two million years after the formation of the Solar System (denoted at CAI formation, the earliest solid material) \citep{Wakita2014}. 

After coalescing from fragments of the disrupted parent asteroid, Itokawa eventually became a near-Earth object. The current orbit suggests that Itokawa originated from the inner part of the asteroid belt, consistent with the prevalence of S-type asteroids observed in this region. Near the belt's inner edge lies the Flora family, a group of asteroids with similar orbital properties that indicate formation from the fragmentation of a common parent body. Members of the Flora family exhibit spectra that is similar to that of Itokawa, and dynamical models support the hypothesis that Itokawa was once a member of this asteroid family \citep{Bottke2025}.

The duration that Itokawa spent in the main belt after formation can be estimated by counting the number of craters visible in the remote-sensing images returned by Hayabusa, and comparing this to models of impact rates within the relatively dense asteroid belt. These estimates range from about 75\,Myr and 1\,Gyr \citep{Michel2009}. Itokawa was most likely delivered into a near-Earth orbit by the $\nu_6$ secular resonance, which is driven by the gravitational influence of Saturn \citep{Michel2006}.

The story of Itokawa therefore began just two million years after the formation of the Solar System with the accretion of a large planetesimal more than 20\,km in radius. Heat from the decay of short-lived radioisotopes raised the internal temperature to around 800\,\degC, causing the minerals to undergo thermal metamorphism while ices and other volatiles were lost. As the planets formed and evolved over the next few billion years, the planetesimal settled into an orbit in the inner asteroid belt, just past 2.0\,au. A catastrophic collision with another asteroid shattered the body into fragments, which coalesced into smaller rubble-pile asteroids. Two of these fragments began to orbit one another, eventually merging to form an elongated shape that resembled a floating otter. Itokawa remained in the main belt for almost 1\,Gyr, gradually drifting due to the Yarkovsky effect from solar radiation until a secular resonance delivered the asteroid into a near-Earth orbit. It is a story that extends from formation over 4.5 billion years ago, through to recent geological changes on the asteroid, all learned from the sample return mission.

\section{The Hayabusa2 mission: sample delivery 2020}

\begin{figure*}[!t]
\centering
\includegraphics[width=\textwidth]{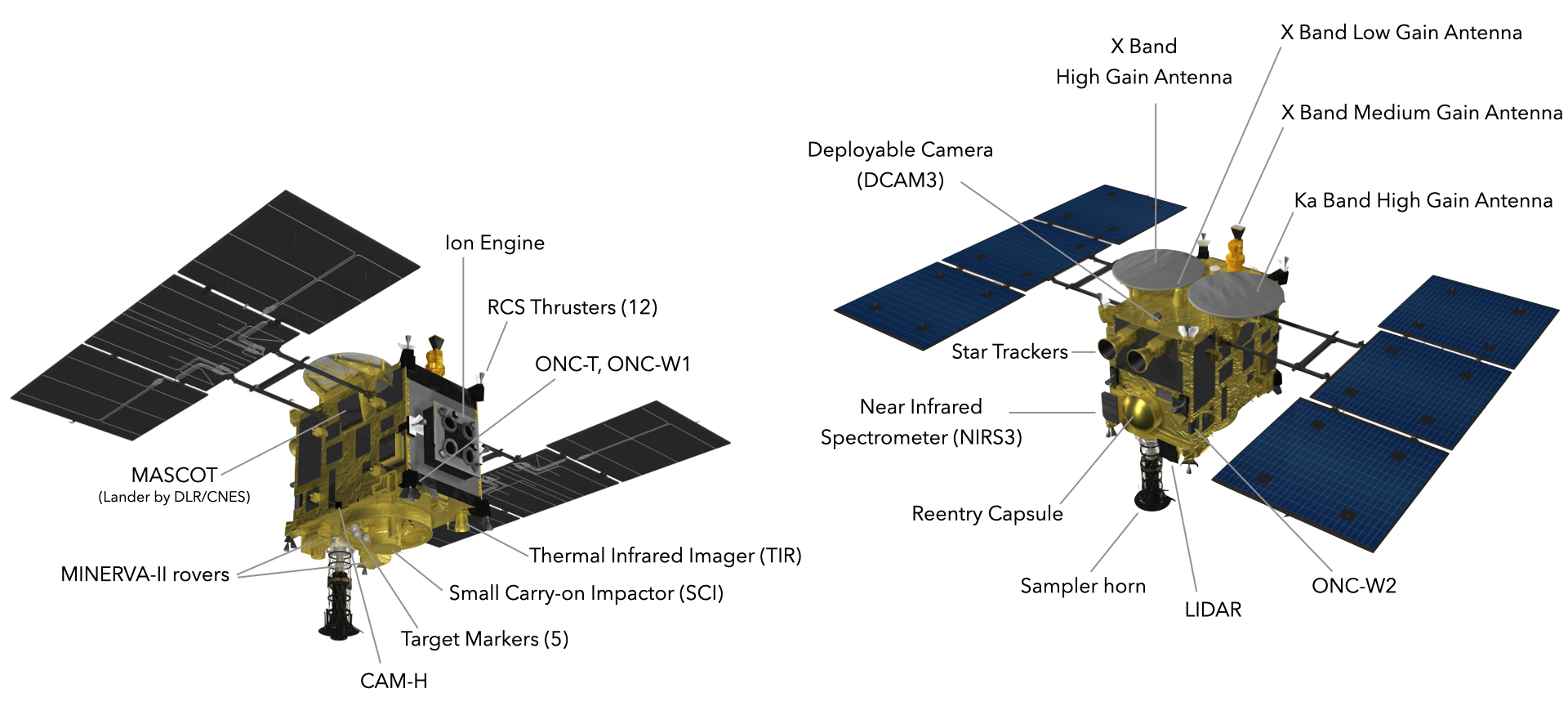}
\caption{Diagram of the Hayabusa2 spacecraft (credit: JAXA).} 
\label{fig:haya2_spacecraft}
\end{figure*}

The second JAXA Hayabusa mission launched from the Tanegashima Space Center on 3 December 2014, and returned to Earth on 6 December 2020. Although the spacecraft design resembled that of its predecessor, the primary objectives of the Hayabusa2 mission were scientific rather than technological. 

Hayabusa2 was to collect and return a sample from the carbonaceous C-type asteroid (162173) Ryugu. In contrast to S-type asteroids such as Itokawa, C-type asteroids were believed to contain pristine material that had not experienced the high temperatures that drive significant thermal alteration. A sample from a C-type asteroid was therefore expected to preserve volatile material such as hydrated minerals and organic matter, offering valuable insights into the distribution of volatiles in the early Solar System and the processes that contributed to the Earth's habitability. 

Launched only four years after the return of Hayabusa, Hayabusa2 inherited the architecture of the previous spacecraft but incorporated a number of improvements informed by the experience of that first mission \citep{Tachibana2014}. The most visible change was the replacement of the single parabolic high-gain antenna with two flat high-gain antennas that enabled high-volume data transmission via Ka-band radio waves (frequency $\sim 32$\,GHz) in addition to the existing X-band ($\sim 8$\,GHz) communications. The ion engines used for interplanetary propulsion also had improved durability and 20\,\% increase in thrust. The damage experienced by Hayabusa led to equipping Hayabusa2 with four reaction wheels (one redundant) for attitude control, and the tip of the sampler horn was also turned up to resemble the teeth of a comb. In the eventuality that the projectiles did not fire to stir surface material during sample collection, the horn tip would provide a passive mechanism to lift pebbles into the sample container. 

The remote-sensing instruments onboard Hayabusa2 include three optical navigation cameras that are used for both spacecraft guidance and scientific observations. These were based on the similar instruments onboard the Hayabusa spacecraft, and consist of two wide-angle cameras (ONC-W1, -W2) and a telescopic camera (ONC-T) equipped with one wideband and seven narrowband filters (0.39 - 0.95\,\micron) for multispectral imaging. The ONC-T has a spatial resolution of 2\,m/pixel at an altitude of 20\,km from the surface, comparable to that of Hayabusa. The instrument suite also includes a near-infrared (1.8 - 3.2\,\micron) spectrometer (NIRS3) for analysing surface mineral composition, a thermal infrared imager (TIR) to measure the thermal emission from the asteroid in the mid-infrared (8 - 12\,\micron), and a laser altimeter (LIDAR) for determining shape and surface topology. An additional camera (CAM-H) with a view alongside the sampler horn was a crowd-funded addition to specifically record the sampling operation (Figure~\ref{fig:haya2_spacecraft}). 

In addition to the remote sensing capabilities, Hayabusa2 carried the Small Carry-on Impact (SCI) experiment, consisting of a 5\,kg copper disc designed to strike the asteroid at approximately 2\,kms$^{-1}$ to generate a new crater and expose subsurface material. To observe the impact, a small wide-angle deployable camera (DCAM3) was included that could detach from Hayabusa2 to record the event while the spacecraft retreated to a safe distance to avoid debris.

The mission also featured surface exploration payloads. The Mobile Asteroid Surface Scout (MASCOT) developed by DLR (German Aerospace Center) and CNES (French National Center for Space Studies) was a compact lander designed to directly analyse the asteroid surface. Three small autonomous rovers called the MIcro Nano Experimental Robot Vehicle for Asteroid (MINERVA-II) were also onboard with the goal to test mobility in a microgravity environment. 

\subsection{The mission}
\label{sec:haya2_mission}

When the Hayabusa2 spacecraft arrived at asteroid Ryugu in June 2018, the first images of the asteroid surface revealed a landscape densely covered with boulders. The journey to Ryugu had proceeded smoothly, but the mission team now faced a significant challenge: the landing sequence used by Hayabusa on asteroid Itokawa would be too dangerous on Ryugu. 

Hayabusa had controlled lateral velocity during touchdown by tracking a falling target marker during the final phase of the descent. However, the uncertainty in the target marker's landing position required a relatively flat area about 100\,m wide. With no sufficiently clear regions on Ryugu, sample collection was postponed from late October 2018 until early the following year to consider how to conduct the operation safely. 

In September and October 2018, the MASCOT lander and two MINERVA-II rovers were successfully deployed, returning images from the asteroid surface that confirmed the rugged terrain. Ryugu was also revealed to be extremely dark, and the sensitivity of the spacecraft LIDAR had to be adjusted to compensate for the low level of reflectivity. 

\begin{figure*}[!th]
\centering
\includegraphics[width=\textwidth]{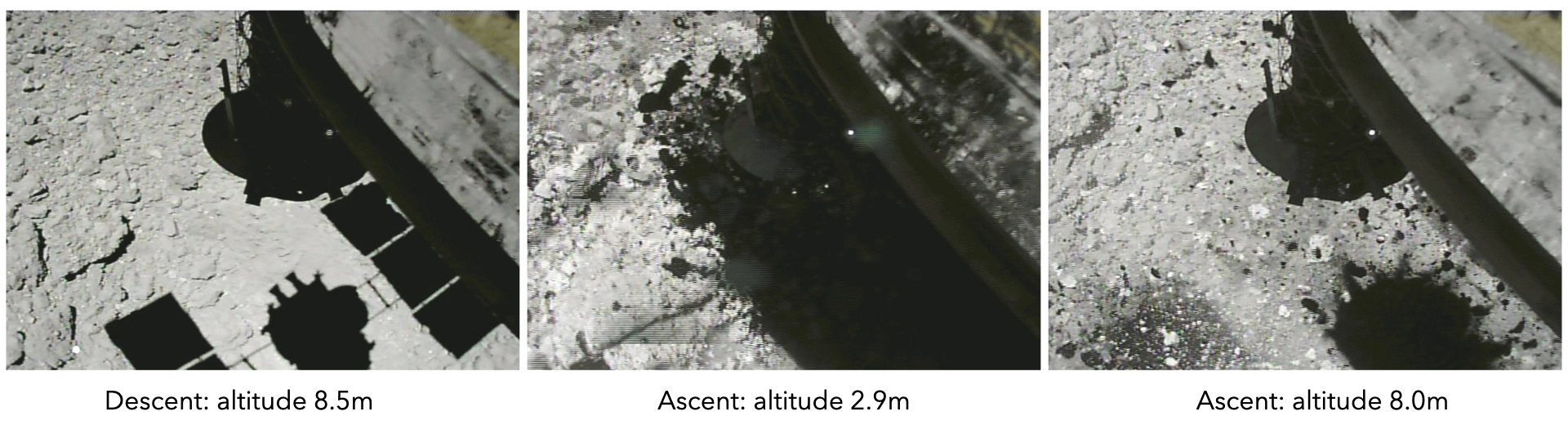}
\caption{Images captured by CAM-H, with a view alongside the sampler horn, during the first touchdown to collect a sample. Touchdown occurred shortly before the middle image was taken, and fragments lifted by the projectile to stir surface material can be seen (credit: JAXA).} 
\label{fig:haya2_camh}
\end{figure*}

On 22 February 2019, Hayabusa2 successfully collected a sample from the asteroid surface (Figure~\ref{fig:haya2_camh}). The new landing technique involved tracking a target marker already placed on the surface and navigating to a position specified relative to this reference point. This method was dubbed `pinpoint landing' and enabled Hayabusa2 to touch down in an area just 6\,m wide with an accuracy of about 1\,m. 

In April 2019, the Small Carry-on Impactor (SCI) experiment was conducted, creating a large crater over 10\,m in diameter. Ejected debris from the impact was scattered across the surrounding region and was identifiable from an even lower albedo than the surface material. Given the hazardous conditions, there was debate over whether to attempt a second sample collection. Ultimately, the scientific value of obtaining subsurface material and the technical challenge justified proceeding. On 13 July 2019, Hayabusa2 made a second successful touchdown in a region about 7\,m wide, located 20\,m from the centre of the artificial crater. The landing accuracy was approximately 60\,cm. 

With the potential for the collected sample to contain hydrated minerals and organics, rigorous contamination control measures were applied throughout the spacecraft development to prevent any compromise from the Earth's environment. Prior to launch, every component of the Hayabusa2 spacecraft that would come into contact with the asteroid underwent a full course cleaning procedure to remove any substances originating on Earth, and the sampler horn was purged with nitrogen gas \citep{Sawada2017}. Samples were collected in the construction areas at Tanegashima Launch Center to record any potential contaminates for comparison during the post-return sample analysis. The sample container itself was hermetically sealed and closed in space after the sample collection was completed. 

Hayabusa2 arrived back at Earth at the end of 2020, and the separated reentry capsule landed in the Woomera Prohibited Area in South Australia on December 6. A sensor on the sample container was designed to change colour if temperatures during atmospheric entry exceeded 65\degC. No colour change was observed, confirming that conditions had remained below daytime temperatures on the asteroid and significantly lower than that needed to dehydrate the sample. The reentry capsule was brought to a `Quick Look Facility' at Woomera, where volatiles were extracted from the sample container to produce the the first gas sample collected from deep space. 

To ensure the integrity of the sample container seal, it was estimated that sample needed to be delivered to the JAXA Extraterrestrial Sample Curation Center within 100 hours after landing. In addition to antennas for detecting the beacon transmitter on the reentry capsule, marine radar stations were deployed to identify the parachute, a drone was used to image the landing area, and the fireball created by the capsule during atmospheric entry was tracked to estimate the landing location. The reentry capsule was located, taken to the Quick Look Facility, and transported back to Japan within 60 hours.  

The sample container was opened in a vacuum clean chamber at JAXA, with the subsequent analysis conducted in clean chambers filled with a high-purity nitrogen. The minimum mission requirement was to return 100\,mg of asteroid material. However, the collected sample weighed 5.4\,g, over fifty times the target amount. The initial characterisation and analysis was performed in Japan, followed by proposals being invited from research laboratories worldwide. 60\% of the sample is not being currently analysed, but has been stored for long-term study.

After the separation of the reentry capsule, the Hayabusa2 spacecraft returned to deep space. The Hayabusa2 Small Hazardous Asteroid Reconnaissance Probe (Hayabusa2$\sharp$) will visit two further asteroids as part of an Extended Mission. 

\subsection{Asteroid Ryugu}

The near-Earth asteroid (162173) Ryugu was first discovered by the LINEAR program in May 1999 and received the provisional designation 1999 JU$_3$. The official name, selected from public suggestions solicited by the JAXA Hayabusa team, refers to the Japanese folktale of Urashima Taro. In this mythological sample return story, the fisherman Urashima Taro visits the undersea palace of Ryugu-jo and brings home a mysterious treasure box. The name `Ryugu' was formally approved by the International Astronomical Union in 2015. 

Like asteroid Itokawa, Ryugu follows an orbit that lies mainly between those of Earth and Mars, ranging between 0.96\,au and 1.41\,au. The asteroid did not pass sufficiently close to the Earth to perform radar observations prior to the arrival of the Hayabusa2 spacecraft, but measurements of the asteroid light-curve revealed smaller variations in the reflected brightness compared to Itokawa. This suggested that Ryugu would have a more spherical shape than the elongated Itokawa.

\subsubsection{A spinning-top asteroid}
\label{sec:haya2_spinningtop}

\begin{figure*}[!th]
\centering
\includegraphics[width=\textwidth]{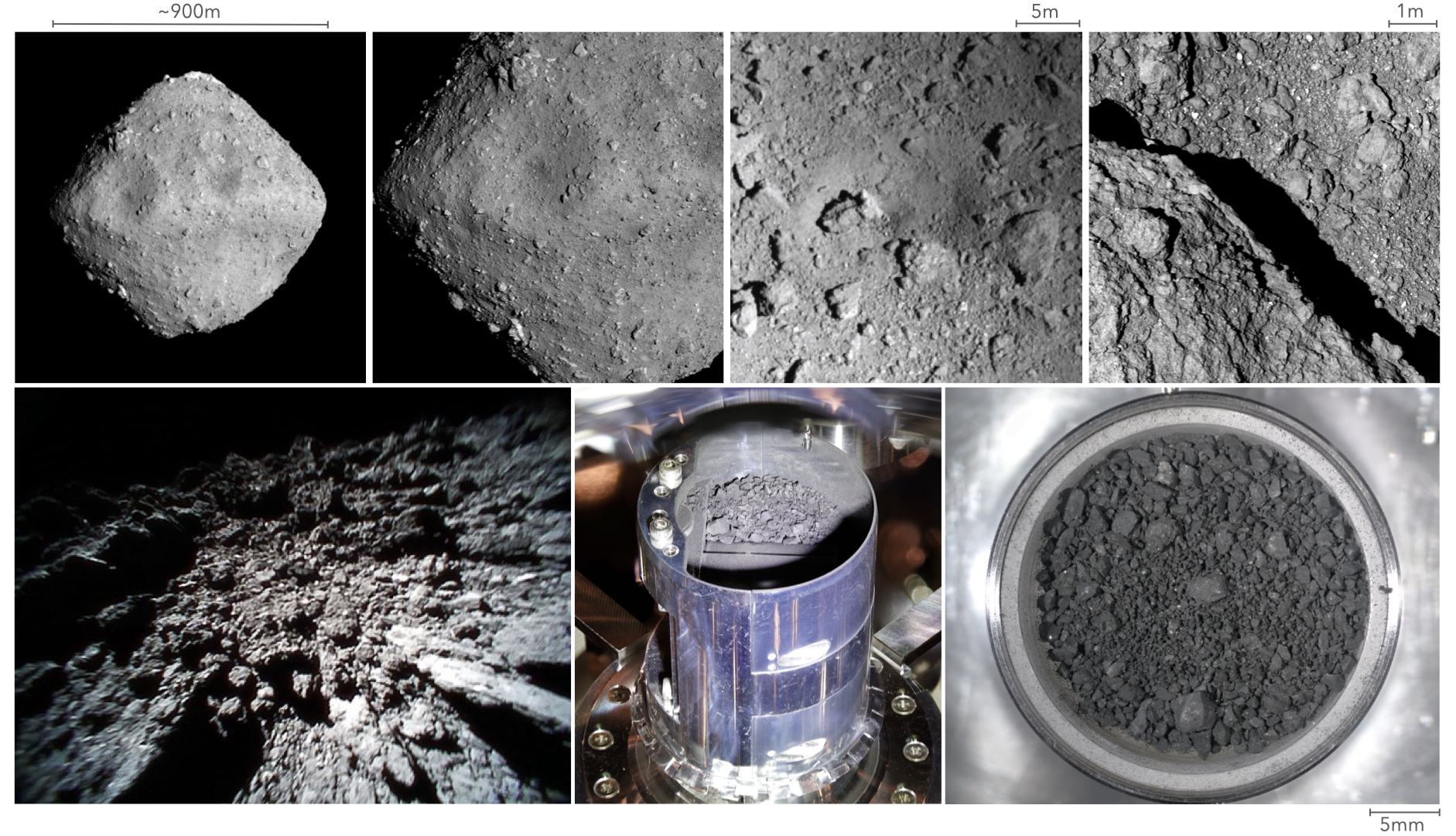}
\caption{Images of asteroid Ryugu and the returned sample. Top row shows images of asteroid Ryugu captured by the Hayabusa2 ONC-T camera. From left to right: asteroid Ryugu from a distance of about 22\,km and 6\,km, the artificial crater made by the SCI impact experiment, and the Ryugu surface from an altitude of about 64\,m during the deployment of two MINERVA-II rovers (ONC-T images: JAXA, University of Tokyo, Kochi University, Rikkyo University, Nagoya University, Chiba Institute of Technology, Meiji University, University of Aizu, AIST). Bottom row (left to right): surface of Ryugu captured by the MINERVA-II (Rover-1B) after landing, the returned sample in chamber A (first touchdown) of the sample catcher within the sample container, optical microscope image of the sample from chamber A (credit: JAXA).} 
\label{fig:ryugu}
\end{figure*}

After arriving at asteroid Ryugu in June 2018, the spacecraft remote-sensing instruments confirmed that the shape of Ryugu is indeed more spherical than that of Itokawa (Figure~\ref{fig:ryugu}). Asteroid Ryugu resembles a `spinning top' with near-conical surfaces towards the poles that meet at a raised circum-equatorial ridge. The radius at the equator measures 502\,m, while the polar-to-equatorial axis ratio is 0.87. A spinning-top morphology can arise from rotational deformation, in which the centrifugal force generated by the asteroid spin pushes material away from the poles and towards the equator. However, the current spin period for Ryugu at 7.6 hours is too slow to induce deformation. Analysis of how the surface slopes on Ryugu (defined as the angle between the surface normal vector and the local total acceleration vector) vary with different spin rates suggests that Ryugu must have once been spinning approximately twice as fast for material to be driven towards the equator and create the current shape \citep{Watanabe2019}. This phase of fast rotation may have occurred during the accumulation of material that formed Ryugu, or it might have been a later gradual change from the YORP effect, where the absorption and re-emission of radiation exerts a net torque. 

\subsubsection{A second rubble pile}
\label{sec:haya2_rubblepile}

The density of asteroid Ryugu was estimated utilising the same method as for asteroid Itokawa. By measuring the gravitational pull from Ryugu during the ballistic descent of the Hayabusa2 spacecraft down to 0.85\,km from the asteroid surface, and subsequent ascent to 5.4\,km, the mass of the asteroid was estimated at $4.50\times 10^{11}$\,kg. Combined with a volume of $3.77 \times 10^8$\,m$^{3}$ calculated from the global shape model for the asteroid generated from images captured by the onboard ONC-T camera, the bulk density of asteroid Ryugu was measured at $1.19$\,\gcm \citep{Watanabe2019}. As with Itokawa, the density of asteroid Ryugu is significantly lower than the density of the delivered sample grains, which have an average bulk density of 1.79\,\gcm. It was therefore possible that asteroid Ryugu shared the loosely-cohesive rubble pile structure of Itokawa.  

Prior to the return of the sample for analysis, it was considered whether the low bulk density of asteroid Ryugu could be explained by the presence of water ice. Remote sensing observations with the Near InfraRed Spectrometer (NIRS3) onboard Hayabusa2 had detected a weak absorption feature centred at 2.72\,\micron across the entire Ryugu surface, indicating the presence of minerals containing the hydroxyl group (an oxygen atom bonded to a hydrogen atom, -OH). However, the radiative equilibrium temperature at Ryugu is approximately 250\,K, and daytime surface temperatures on the asteroid measured by the Thermal Infrared Imager (TIR) reached up to 370\,K. These values are higher than the ice sublimation temperature at about 230\,K, even at the estimated central pressure within the asteroid ($\sim 8$\,Pa). The low density of the asteroid must therefore be due to porosity, not light ices \citep{Watanabe2019}.

Additional confirmation of a rubble pile structure came from the visual inspection of the asteroid surface. Unlike the regions of smooth terrain present on asteroid Itokawa, the surface of Ryugu is homogenously covered with boulders, as can be seen in the surface images in Figure~\ref{fig:ryugu}. (This made the touchdown of the spacecraft particularly challenging, as described in section~\ref{sec:haya2_mission}.) The global number density of boulders exceeding 20\,m in diameter on Ryugu is more than twice that of Itokawa \citep{Watanabe2019, Sugita2019}.  Moreover, even the 290\,m diameter Urashima Crater on Ryugu (the largest on the asteroid surface) is not sufficiently large to have ejecta that could have produced the biggest ($>20$\,m) boulders. The boulders are therefore likely fragments of a parent body that catastrophically disrupted and loosely re-accreted to form Ryugu \citep{Watanabe2019}.

The boulders themselves were also discovered to be surprisingly porous. Images taken by the TIR showed that the majority of the boulders on Ryugu were at a similar temperature to their surroundings. Boulders are typically expected to be dense structures with high thermal inertia, changing temperature more slowly than the surrounding material. Dense boulders would therefore have been seen as cold spots on the daytime side of the asteroid. This was observed for a small number of boulders seen on Ryugu, which have derived thermal inertia values between $600 - 1000$\,Jm$^{-2}$s$^{-0.5}$K$^{-1}$ ($600 - 1000$\,tiu). However, the  population is dominated by boulders with a thermal inertia of only 300\,tiu, suggesting that the rock is generally highly porous. This conclusion was supported by the MASCOT lander, which returned images of boulders with crumbly surfaces suggestive of a low density, and a measured thermal inertia for one boulder of 282\,tiu that was in good agreement with the TIR data. These findings imply that the parent body of Ryugu was also made from porous material, poorly consolidated due to low gravity conditions except for a fraction of high density rock forming in the deepest part of the asteroid. The high thermal inertia of the rarer boulders is consistent with that of carbonaceous chondrites, suggestive of a selection effect whereby the principal fluffy material that constitutes C-type asteroids is destroyed before reaching the Earth surface. Accounting for the high porosity of the boulder population, asteroid Ryugu is estimated to have a macroporosity of about 20\,\%. The asteroid is therefore a loosely packed rubble pile, made from porous material \citep{Okada2020, Grott2020}.

\subsubsection{The link between C-type asteroids and carbonaceous chondrites}
\label{sec:haya2_CIchondrite}

\begin{figure*}[!th]
\centering
\includegraphics[width=\textwidth]
{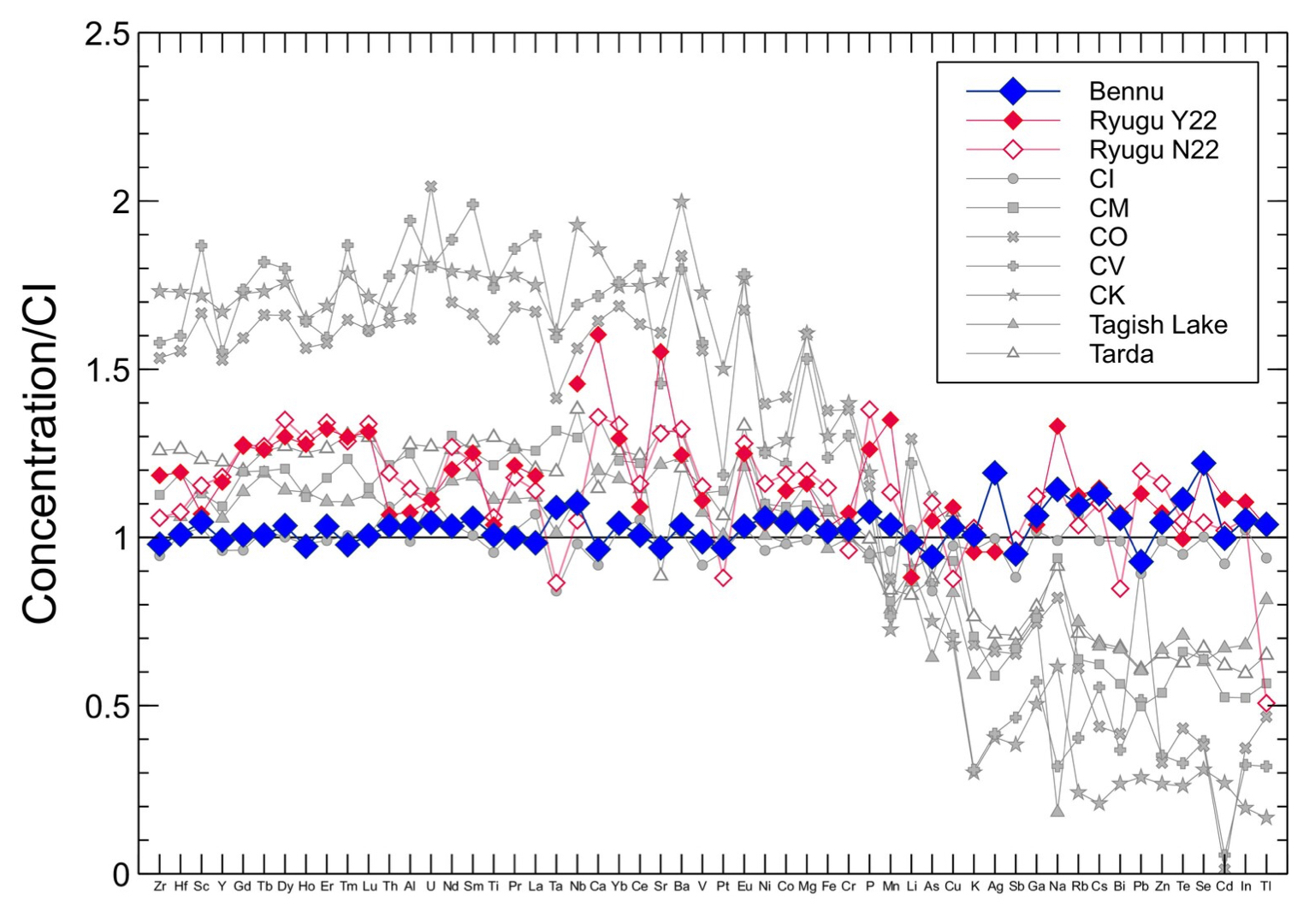}
\caption{Elemental abundances listed by increasing volatility (decreasing 50\% condensation temperature) in the bulk Ryugu returned sample reported by \citet{Yokoyama2022} (Y22) and \citet{Nakamura2022} (N22), and the Bennu returned sample, in comparison to different carbonaceous meteorite types, all normalised to CI chondrite abundances (reproduced from Lauretta \& Connolly et~al.~(\citeyear{LaurettaConnolly2024}) licensed under CC BY 4.0).} 
\label{fig:ryugu_elements}
\end{figure*}

During the approach to asteroid Ryugu, the Hayabusa2 spacecraft measured the disc-averaged colour of the asteroid through each of the ONC-T seven broadband filters. The resulting spectrum confirmed the ground-based classification that Ryugu is a C-type (carbonaceous) asteroid. The spectrum and orbit of Ryugu suggest that the asteroid may be a fragment of the new Polana family, a group consisting of C- and B-type asteroids in the inner main belt suspected to have formed in a major disruption event approximately 1500 million years ago. In contract, the surface age of Ryugu based on crater count and retention estimates is only about 8.9 million years, indicating that the asteroid might be the product of a later generation of collision and re-accumulation rather than a direct product of the initial breakup \citep{Arakawa2020, Sugita2019}.

Initial laboratory analysis of approximately 125\,mg of material returned from the two touchdown sites measured the relative abundances of 66 chemical elements from hydrogen through to uranium \citep{Yokoyama2022}. Except for the tantalum introduced by the sampling projectiles, the chemical composition of the Ryugu sample closely matches that of CI chondrite meteorites, with a notably high abundance of volatile elements compared to the other carbonaceous chondrite groups (Figure~\ref{fig:ryugu_elements}). No systematic difference was found between samples from the two touchdown sites. Isotopic ratios for elements including titanium, chromium, oxygen, potassium, zinc, copper, and hydrogen also strongly align with CI chondrites \citep{Hu2024, Paquet2023, Piani2023}.  

The composition of the Ryugu sample was not identical in all respects to that of CI chondrites. The resulting minerals in the Ryugu sample lack sulphates and ferrihydrite, and contain substantially less water trapped between the layers of the hydrous silicate minerals (phyllosilicates). These discrepancies likely result from terrestrial alteration of the meteorites after landing on Earth. The interlayer water may derive from atmospheric moisture, while both sulphates and ferrihydrite are products of weathering reactions with oxygen and water. A small offset in the abundance of the oxygen isotope $^{17}$O further supports the explanation that the meteorites were affected by water on Earth which led to both structural and chemical changes. The sample from Ryugu therefore consists of a more chemically pristine form of CI chondrite material \citep{Yokoyama2022}.

As discussed in section~\ref{sec:meteorites}, carbonaceous chondrites, and especially CI chondrites, are the rarest class of meteorites. By contrast, ground-based observations indicate that C-type asteroids dominate in the Solar System, comprising of more than 75\% of known asteroids. If the majority of C-type asteroids are carbonaceous chondrites similar to Ryugu, then this discrepancy points to a strong sampling bias in the meteorite collection. The probable cause is the fragility of carbonaceous chondrite material, which is rarely surviving the intense aerodynamic stresses of atmospheric entry. This is consistent with the high porosity of Ryugu observed in thermal images (section~\ref{sec:haya2_rubblepile}) and the mechanical properties of the delivered sample grains, which were substantially less hard than terrestrial igneous rocks and could be cut with a knife, and also displayed fine cracks \citep{Nakamura2022}. Images from the Hayabusa2 small monitor camera (CAM-H) during the first touchdown operation also captured a centimetre-sized pebble fragmenting after a low-velocity impact of about 0.1\,ms$^{-1}$. This suggests either a very low tensile strength or pervasive internal fractures \citep{Tachibana2022}.

CI chondrites are considered the closest match to the solar photosphere and the initial composition of the Solar System. The sample from Ryugu is therefore the most pristine example yet obtained of the material that would have formed the planets. The fragility of the asteroid also explains why such material is under-represented amongst meteorites, despite the abundance in the asteroid population.

\subsubsection{Aqueous alteration}
\label{sec:haya2_aqueous}

\begin{figure*}[!th]
\centering
\includegraphics[width=\textwidth]{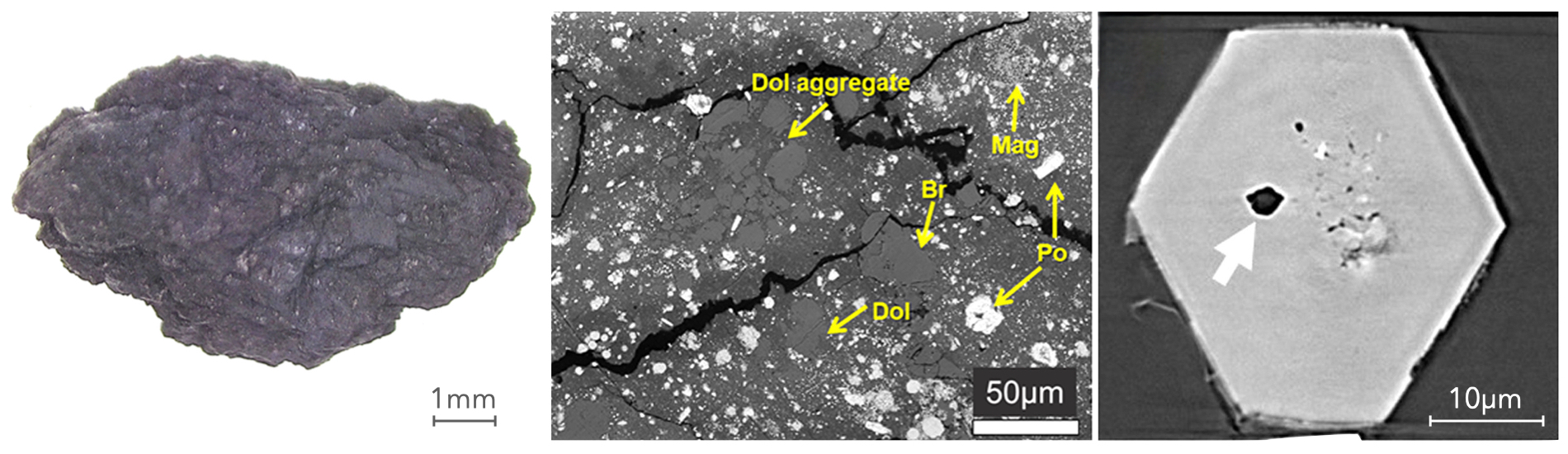}
\caption{Morphology, texture, and fluid inclusions in the Ryugu sample grain, C0002. From left to right: optical microscope image of the whole sample grain, electron microscope (back-scattered electron) image of the typical internal texture, labelling regions of dolomite (Dol), magnesite (Br), pyrrhotite (Po) and magnetite (Mag) embedded within the phyllosilicate matrix, slice through the SR-nanoCT scan showing the fluid inclusion, indicated by the arrow (reproduced with permission from \citet{Nakamura2022} \raisebox{0.3ex}{\scalebox{0.8}{\copyright}}\,2023 AAAS).} 
\label{fig:ryugu_fluid}
\end{figure*}

Although the chemical composition of the Ryugu sample closely matches that of the solar photosphere, the mineralogy does not consist of primary materials that solidify directly from the protosolar nebula. The delivered grains predominantly comprise of secondary minerals that are formed through interaction with liquid water, a process known as `aqueous alteration'. The principal sample constituents are phyllosilicates, which are silicate minerals with a characteristic layered structure. Phyllosilicates are typically hydrated minerals, incorporating either OH hydroxyl groups (-OH) or water molecules (H$_2$O) into their structure. In the Ryugu sample, most of the water is hosted by two principal phyllosilicates: serpentine with the chemical formula (Mg,Fe)$_3$Si$_2$O$_5$(OH)$_4$, and saponite with chemical formula Ca$_{0.25}$(Mg,Fe)$_3$((Si,Al)$_4$O$_{10}$)(OH)$_{2n}$(H$_2$O). Other major minerals include carbonates such as dolomite (CaMg(CO$_3$)$_2$) and iron- and manganese-bearing magnesite ((Mg,Fe,Mn)CO$_3$), iron sulphides like pyrrhotite (Fe$_{1-x}$S), and magnetite (Fe$_3$O$_4$), all of which form through precipitation from aqueous solutions (see Table~\ref{table:ryugu_minerals} and Figure~\ref{fig:ryugu_fluid}). As with the chemical element abundance, this mineral assemblage is a close match to that found in CI chondrites. 

\begin{table*}[t]
\caption{Abundance (by volume percentage, vol\%) of the main minerals in the
largest Ryugu sample grain \citep{Nakamura2022}. Detailed mineralogy of the Ryugu
samples are described in \citet{Nakamura2022, NakamuraEizo2022, Nakato2022, Noguchi2024}.}
\label{table:ryugu_minerals}

\vspace{4pt}

\centering
\begin{tabular*}{\linewidth}{@{\extracolsep{\fill}} l l r}
\hline
Mineral group & Formula & vol\% \\
\hline
Phyllosilicates &  & 86.2 \\
\hspace{3mm}Serpentine & (Mg,Fe)$_3$Si$_2$O$_5$(OH)$_4$ &  \\
\hspace{3mm}Saponite & Ca$_{0.25}$(Mg,Fe)$_3$((Si,Al)$_4$O$_{10}$)(OH)$_{2n}$(H$_2$O) &  \\[1ex]

Magnetite & Fe$_3$O$_4$ & 5.7 \\[1ex]

Fe sulphides &  & 4.2 \\
\hspace{3mm}Pyrrhotite & (Fe,Ni)$_{1-x}$S &  \\
\hspace{3mm}Pentlandite & (Fe,Ni)$_9$S$_8$ &  \\[1ex]

Dolomite & CaMg(CO$_3$)$_2$ & 2.4 \\[1ex]
Magnesite & (Mg,Fe,Mn)CO$_3$ & 1.2 \\[1ex]
\hline
\end{tabular*}
\end{table*}

Because of Ryugu's small size and resulting low internal pressure, liquid water would not be stable within the present asteroid interior. Aqueous alteration must therefore have occurred at an earlier epoch on the larger parent body. The timing of this process can be estimated by dating the formation of the mineral dolomite using $^{53}$Mn - $^{53}$Cr chronometry. The chromium isotope, $^{53}$Cr, is the daughter product of the decay of the short-lived radioactive manganese isotope $^{53}$Mn, which has a half-life of 3.7 million years. The excess of $^{53}$Cr in the Ryugu sample can be used to estimate the time since dolomite first formed. The result indicate that the minerals of Ryugu formed only about four to five million years after the onset of the Solar System formation \citep{Yokoyama2022, Sugawara2024}.

This early formation is further supported by measurements of remanent magnetism discovered in 300 - 110\,nm aggregates of magnetite. It is unlikely that this magnetism originated from the parent body of Ryugu, as the compositional homogeneity seen in both the remote sensing data and the returned sample suggests that the parent was not a differentiated body. The magnetic field is therefore likely to be recording that of the solar nebula that surrounded the young Sun. This implies that the magnetite formed during the earliest stages of planet formation, before the solar nebula dispersed; it now effectively preserves a magnetic `recording' similar to a hard disc, but approximately 4.6 billion year old \citep{Nakamura2022}.

The temperature within the Ryugu parent body during the aqueous alteration can be estimated by the isotopic ratios of oxygen in the precipitated dolomite and magnetite. Depending on the temperature of the solution, the two minerals acquire varying relative abundances of the heavier $^{18}$O isotope. By comparing the difference in $^{18}$O content between dolomite and magnetite and applying an experimentally derived temperature calibration, the temperature at precipitation can be determined. This technique is known as oxygen isotope thermometry. Independent analyses estimate that the interior of Ryugu during aqueous alteration had temperatures between approximately 40\degC to 80\degC  \citep{Yokoyama2022, NakamuraEizo2022, Kita2024}. 

Although the Ryugu sample lacks significant levels of interlayer water (section~\ref{sec:haya2_CIchondrite}), a minor signature of interlayer water was detected when the sample was heated to 90\degC. This suggested that even since aqueous alternation, the Ryugu sample has never experienced temperatures above 100\degC, confirming its chemically pristine nature \citep{Yokoyama2022}.

In order to have liquid water, the parent body of Ryugu most likely formed beyond the ice line in the outer Solar System, where temperatures were sufficiently low for ice to condense and be accreted as the asteroid formed. Evidence for this distant formation came from the discovery of trapped fluid inclusions in a large pyrrhotite (iron sulphide) crystal (right-hand image in Figure~\ref{fig:ryugu_fluid})\citep{Nakamura2022}. The fluid was found to contain both water and carbon dioxide, indicating that conditions were cold enough for both these molecules to exist as solid ice. Even accounting for the reduced solar luminosity in the early Solar System, the ice lines would have been beyond 3 - 4\,au from the Sun. Despite forming in the outer Solar System, Ryugu does not appear to be the fragment of a comet. The carbon-to-silicon ratio in the Ryugu sample (C/Si = 0.338), is lower than that of comets, based on measurements of interplanetary dust particles that have a probable cometary origin. 

\subsubsection{Organics}
\label{sec:haya2_organics}

When the Hayabusa2 spacecraft arrived at asteroid Ryugu, it found a surface darker than any celestial object previously visited. At a wavelength of 0.55\,\micron, remote-sensing observations with the ONC-T camera measured a geometric albedo for Ryugu of just 4.5\%. The composition of Ryugu therefore must include a highly absorbing component, with a likely candidate being a high abundance of carbon \citep{Sugita2019, Kitazato2019}.  

Analysis of the delivered sample revealed that Ryugu is indeed about $4$\,\% carbon, with approximately two-thirds of this carbon present in organic molecules \citep{Yokoyama2022, Nakamura2022}. Organic molecules are typically defined to be those containing carbon bonded either to hydrogen or another carbon atom. These compounds form the molecular components essential for terrestrial life, including amino acids, enzymes, the double- and single-stranded nucleic acids DNA and RNA, sugars, fatty acids, and cell membranes. The emergence of organic molecules is therefore of great interest for investigating the origins of our planet. Organic molecules have been found meteorites, particularly in carbonaceous chondrites that have undergone aqueous alteration where the liquid water medium can promote the synthesis of new and more complex molecules. This suggests that these building blocks for terrestrial life may have originally formed in small celestial bodies and were subsequently delivered to Earth through meteorite impacts.  

More than 20,000 distinct organic molecules have been identified in the Ryugu sample, including approximately 20 amino acids and all five primary nucleobases (adenine, guanine, cytosine, thymine, and uracil) that form the base sequences of RNA and DNA \citep{Koga2026, Naraoka2023, Oba2023, NakamuraEizo2022}. Proteinogenic amino acids (those used by living organisms to form proteins) such as glycine, alanine, and valine were detected alongside non-proteinogenic amino acids. The presence of proteinogenic amino acids in a sample isolated from Earth's biosphere confirms that these compounds can be delivered in extraterrestrial material and are not solely the result of contamination in meteorites. The organics and phyllosilicates within the sample are closely associated, suggesting that aqueous alteration and the diversification of organic molecules evolved together on the Ryugu parent body \citep{Yabuta2023}. Given that the measured epoch for the aqueous alteration was only a few million years after the formation of the Solar System, this implies that the first building blocks of life are already present in early planetary systems and could have been delivered to the early Earth.

With the exception of glycine, the naturally occurring amino acids on Earth are chiral, possessing two mirror-image molecular structures that cannot be superimposed. Terrestrial life exclusively uses the left-handed (L-) form of amino acids and not the right-handed (D-) form. The reason for this preference remains unknown, and a related outstanding question is whether the bias originated on Earth during the genesis of life or if extraterrestrial material delivered a biased supply of organics to the early Earth. In the Ryugu sample, both D- and L- enantiomers exist in approximately equal quantities. This confirms that the amino acids are of extraterrestrial origin and suggests that the initial supply of organics to the young Earth was a racemic mixture. 

\begin{figure*}[!th]
\centering
\includegraphics[width=\textwidth]{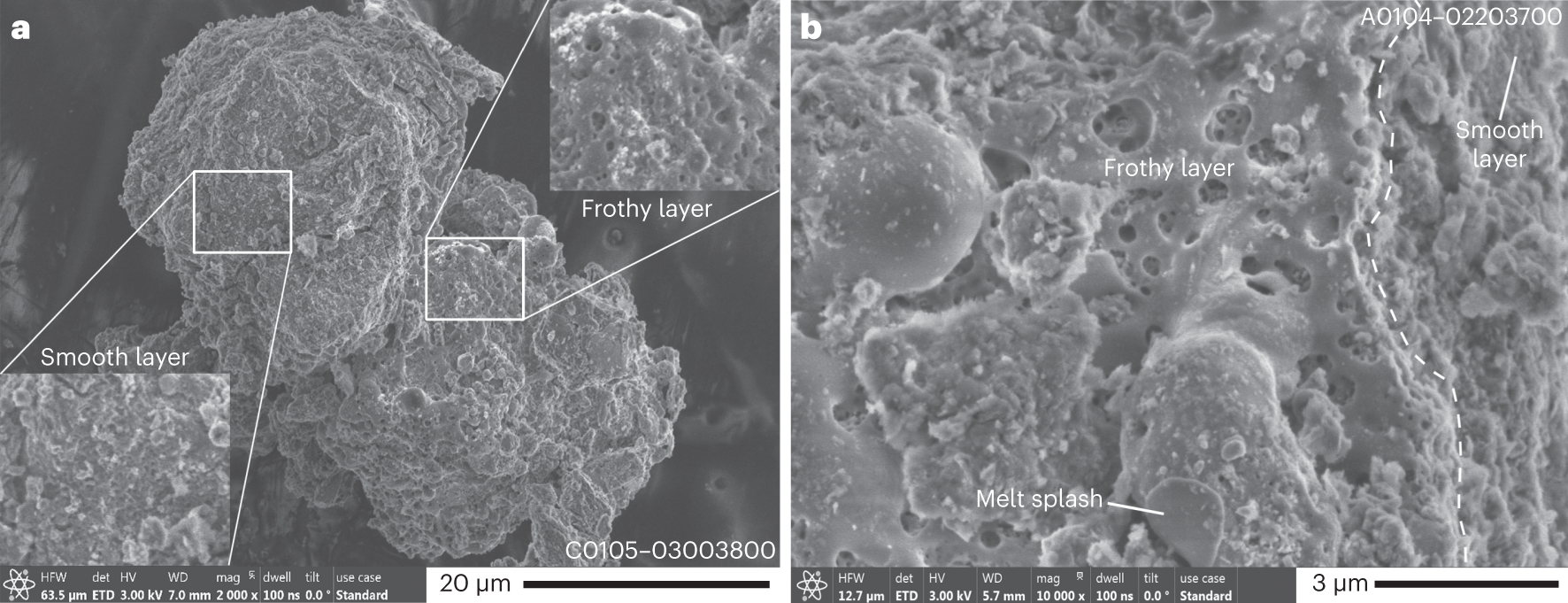}
\caption{Electron microscope image (secondary electron image) of different types of space weathering visible on the Ryugu sample grains: a smooth layer created by the solar wind and a frothy layer from micrometeorite impacts (reproduced from \citet{Noguchi2023} licensed under CC BY 4.0).} 
\label{fig:ryugu_spaceweathering}
\end{figure*}

\subsubsection{Presolar stardust}
\label{sec:haya2_presolar}

While the majority of the Ryugu sample exhibits a uniform isotopic composition, a small fraction of sub-micrometre grains show anomalous enrichment or depletion of certain isotopes, suggesting an origin independent of the bulk asteroid matrix. Similar inclusions have previously been detected in primitive meteorites and are considered to be presolar grains that condensed in the outflows of evolved stars or in supernovae ejecta. These grains survived incorporation into the solar nebula, and subsequent processing within the asteroid. Presolar grains with anomalous carbon or oxygen isotopic signatures were identified in the Ryugu sample in abundances and compositions consistent with CI chondrites \citep{Barosch2022}. 

Complex organic molecules in the form of polycyclic aromatic hydrocarbons (PAHs) have also been identified in the Ryugu sample, with a similarly suspected origin from predating the formation of the Sun. PAHs are ubiquitous in the interstellar medium (ISM) and are estimated to contain approximately 20\,\% of ISM carbon atoms. Like presolar grains, PAHs have also been found in primitive meteorites, including CI chondrites. The principal formation location of PAHs is debated. However, analysis of the PAHs found in the Ryugu sample revealed an enhanced yield of chemical bonds between two $^{13}$C atoms; which is the heavier, less abundant carbon isotope. This pattern suggests formation at low temperatures, where isotope fractionation occurs due to the stronger bond strength of the heavier atoms becoming significant when thermal energy is limited. It is therefore likely that the PAHs in the Ryugu sample formed within the cold (approximately 10\,K) molecular clouds within the ISM \citep{Zeichner2023}. Organic material in the sample that was identified with elevated abundances of deuterium and $^{15}$N may also share this origin, forming under conditions where heavier isotopes are favoured in the cold molecular clouds \citep{Yabuta2023}.

The discovery of presolar grains, and ISM PAHs, and $^{15}$N- and deuterium-rich organics further reinforces the connection between Ryugu and CI chondrites, and also demonstrates that the Earth likely acquired a portion of solid material and organics from beyond the Solar System.

\subsubsection{A cloaked asteroid}
\label{sec:haya2_spaceweathering}

Shortly after arrival at Ryugu, the Hayabusa2 NIRS3 infrared spectrometer conducted a near-global scan of the asteroid surface. An absorption feature centred at 2.72\,\micron was detected across the entire observed region, indicating the ubiquity of hydrated minerals carrying the hydroxyl (-OH) group. However, the absorption feature was weak, with an intensity only $7-10$\,\% below the spectral continuum (the baseline spectrum in the absence of absorption features). The weakness of the feature suggested that while Ryugu had undergone aqueous alteration, the asteroid had either experienced significant heat and become dehydrated, potentially by passing close to the Sun, or that the water content in the parent body had been relatively low \citep{Kitazato2019}. From the perspective of investigating an early mineralogy capable of supporting prebiotic chemistry, this might have been disappointing. However, it would later prove to be in sharp contrast to the extensive aqueous alteration revealed during the analysis of the delivered sample (section~\ref{sec:haya2_aqueous}). So why did the remote-sensing observations and sample analysis differ? 

As with asteroid Itokawa, variations in the reflected spectra across the asteroid surface provided an early indication that the discrepancy might be due to space weathering (see section~\ref{sec:haya_spaceweathering}). Observations with the Hayabusa2 ONC-T camera revealed that the interior of stratigraphically younger craters, such as those with less eroded rims and superimposed over older craters, with diameters greater than 20\,m have interiors that are spectrally bluer that the surrounding surface \citep{Morota2020}. This suggests that the buried subsurface material of Ryugu had a slightly different composition to the asteroid uppermost layer. A similar variation was also observed as a global trend, with the prominent equatorial ring and polar regions appearing spectrally bluer than areas at mid-latitude. This distribution is consistent with mass wasting, whereby material from the equator and poles migrated into the mid-latitude regions as the asteroid spin rate changed (section~\ref{sec:haya2_spinningtop}), exposing bluer material beneath the displaced top layer.

\begin{figure*}[!th]
\centering
\includegraphics[width=\textwidth]{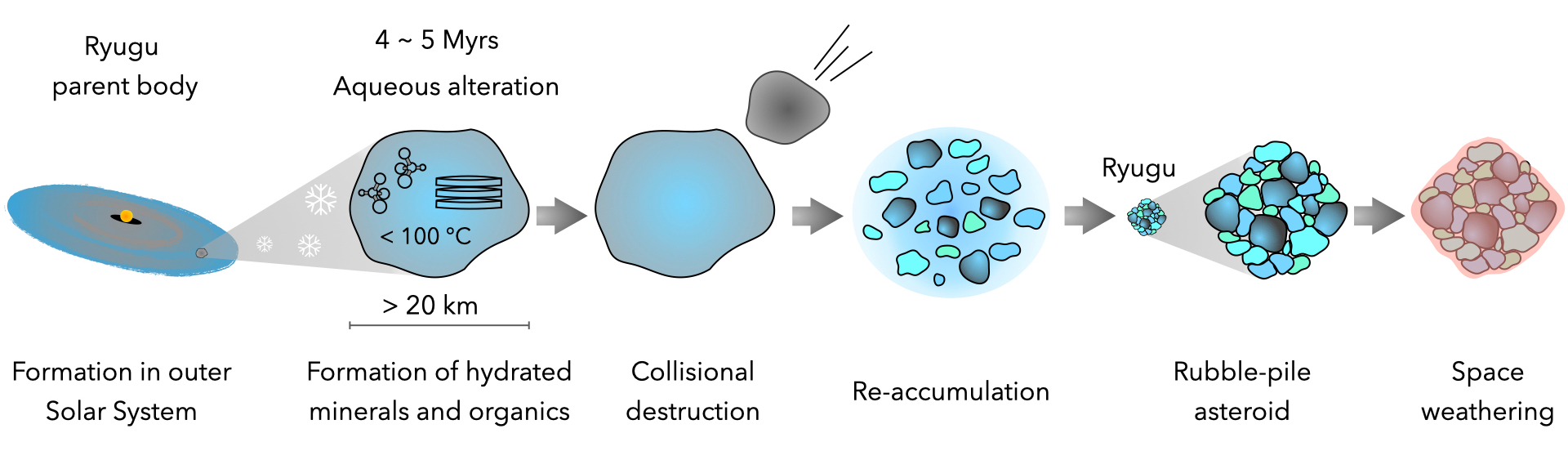}
\caption{The history of asteroid Ryugu.} 
\label{fig:ryugu_history}
\end{figure*}

Direct evidence that Ryugu had been affected by space weathering came from the  gas trapped within the returned sample container. The gas consisted of helium and neon with isotopic ratios that differed from those of the Earth's atmosphere and matched those of the solar wind. This was therefore gas that was implanted in the Ryugu grains during bombardment by the solar wind and subsequently released as the grain surfaces were pulverised during collection \citep{OkazakiGasSample2022}. 

Confirmation that space weathering is driving dehydration in the uppermost layer of Ryugu was found in the returned sample. Detailed examination of the Ryugu grains revealed distinct changes to the phyllosilicate minerals in about 10\,\% of the grains. The affected grains were dehydrated, with either a smooth or frothy outer layer (Figure~\ref{fig:ryugu_spaceweathering}). Laboratory experiments designed to mimic space weathering suggested that solar wind irradiation leaves a smooth layer on phyllosilicate material, while bombardment with micrometeorites gives a frothy layer with evidence of melt splashes. Both effects cause dehydration, with the matrix stripped of the structural -OH hydroxyl group and the material taking on a redder spectral hue. This process is different from space weathering observed on the Moon and Itokawa (section~\ref{sec:haya_spaceweathering}), due to the hydrated layered silicates responding differently to anhydrous silicates \citep{Noguchi2023}.

This examination confirmed that strong thermal alteration had not occurred throughout the asteroid parent body, since the majority of the phyllosilicates preserved their hydroxyl groups. Furthermore, the observed dehydration is most consistent with space weathering, rather than a sunburn from intense solar radiation, as the latter would have produced more pervasive effects and strongly heated grains at greater depths. This additionally indicates limited mixing between the surface and material only a few metres down. Such agitation is typically driven by meteorite impacts, whose frequency would have dropped considerably once Ryugu left the asteroid belt and entered a near-Earth orbit. Combined with crater counts, this supports a conclusion that Ryugu migrated from the asteroid belt to its current orbit about 5 million years ago \citep{Okazaki2022}. 

Suppression of the 2.7\micron absorption band by space weathering has significant implications for ground-based observations of asteroids. Asteroids whose spectra indicate a lack of water may possess only a dry outer layer. This will need to be taken into consideration when interpreting the distribution of water in space. 

\subsubsection{The history of Ryugu, and a path to life on Earth}

Four to five million years after the beginning of the Solar System, solid material condensing from the gaseous solar nebula aggregated into a loosely bound planetesimal. Presolar grains and organic compounds that had originated outside the Solar System before the formation of the Sun were also incorporated into this primitive body, which grew to at least 20\,km in diameter. This was the parent body of asteroid Ryugu (Figure~\ref{fig:ryugu_history}).

Formation most likely occurred in the outer region of the Solar System, at least $3-4$\,au from a cooler young Sun. Situated beyond the ice lines for both water and carbon dioxide, these volatiles were accreted into the planetesimal as solid ice. A reduced abundance of the the short-lived radionuclide $^{26}$Al compared to that in the earlier-forming parent body of asteroid Itokawa prevented intense thermal metamorphism. Instead, the internal heat melted ice into liquid water with temperatures around $40 \sim 80$\degC. The resulting extensive aqueous alteration transformed the minerals and triggered the formation of tens of thousands of organic molecules, including a racemic mixture of amino acids and uracil. With the planetesimal still embedded in the solar nebula, crystals of magnetite precipitating from the liquid recorded the nebula magnetic field.

As planet formation progressed, the planetesimal was scattered inward towards the inner asteroid belt, where a major impact disrupted the body. Asteroid Ryugu coalesced as a rubble pile from the resulting debris, forming either directly from the disrupted parent planetesimal or after the subsequent fragmentation of an intermediate parent. About 5 million years ago, Ryugu migrated into a near-Earth orbit. Over the next few millennia, the surface of Ryugu reddened through space weathering, hiding a water-rich interior that preserved all but the most highly volatile original ingredients of the Solar System. 

Other fragments of the parent body, and planetesimals with similar compositions, could have collided with the young terrestrial planets. If the inventory of asteroid Ruygu is typical, this would have delivered water-rich minerals and organic compounds that may have contributed to the emergence of habitability. But is asteroid Ryugu typical of primitive material or unusual? The only way to know is to compare with more samples.

\section{The OSIRIS-REx mission: sample delivery 2023}

The first asteroid sample return mission led by NASA launched on 8 September 2016. The Origins, Spectral Interpretation, Resource Identification, and Security-Regolith Explorer (OSIRIS-REx) mission conducted a seven-year round trip journey to the near-Earth Object (101955) Bennu, delivering a sample to Earth on 24 September 2023. 

Like asteroid Ryugu, ground-based observations of Bennu identified the asteroid as carbonaceous, but a relatively blue spectrum classified Bennu as a B-type asteroid. With the hope that Bennu was related to carbonaceous chondrite meteorites, the primary objective of the OSIRIS-REx mission was to deliver a pristine sample whose analysis could provide information on the source of the Earth's water and organic material. 

An additional mission objective related to the presence of Bennu in a near-Earth orbit. The asteroid is considered one of the most potentially hazardous known celestial objects, with a small chance of an Earth impact in the second half of the next century. The occurrence of impact depends on the precise future trajectory, whose uncertainty is dominated by non-gravitational perturbations such as the Yarkovsky effect. The mission would therefore study the Yarkovsky effect on Bennu to improve both the long-term predictions for the asteroid's position, and also to better understand the orbital evolution of similar small bodies with Earth-crossing orbits. 

\begin{figure*}[!th]
\centering
\includegraphics[width=\textwidth]{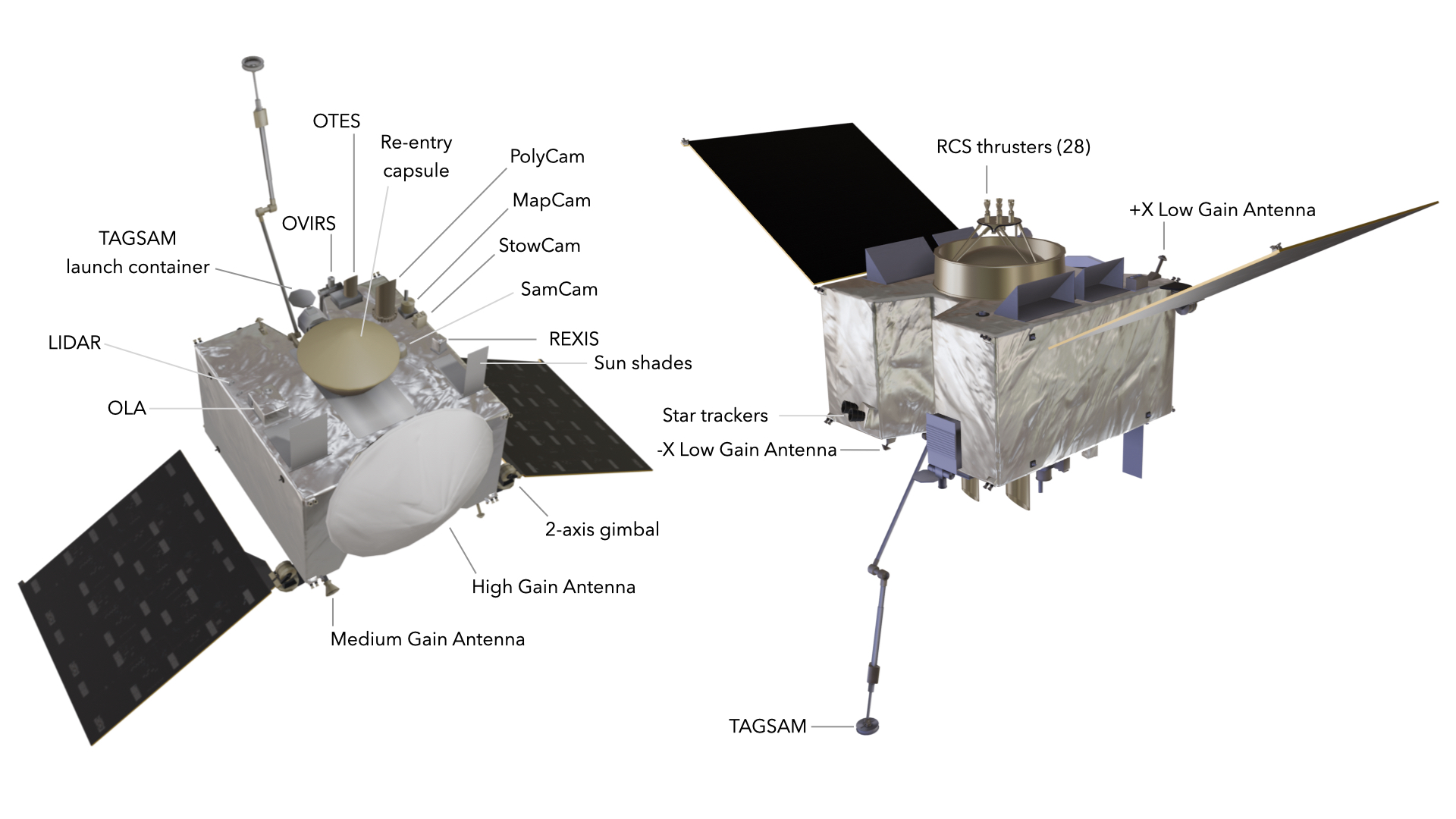}
\caption{Diagram of the OSIRIS-REx spacecraft (spacecraft model: NASA VTAD).} 
\label{fig:orex_spacecraft}
\end{figure*}

The OSIRIS-REx spacecraft was equipped with five instruments for the remote-sensing part of the mission (Figure~\ref{fig:orex_spacecraft}). The OSIRIS-REx camera suite (OCAMS) consists of three cameras for navigation and capturing images of the asteroid surface: the telescopic PolyCam for long-range observations and high resolution imaging (better than 30\,cm/pixel at an altitude of 20\,km from Bennu); the medium-resolution MapCam for surface mapping and satellite or plume detection, equipped with four filters spanning visible and near-infrared wavelength band; and the wide-angle SamCam for observing the sampling operation. OSIRIS-REx also carried three spectrometers: the OSIRIS-REx Thermal Emission Spectrometer (OTES) for mid-infrared (5 - 100\,\micron) thermal measurements; the Visible and Infrared Spectrometer (OVIRS) covering visible to infrared (0.4 - 4.3\,\micron) to identify mineral composition; and the student-led Regolith X-ray Imaging Spectrometer (REXIS) experiment for elemental analysis via X-ray induced fluorescence. Finally, the Canadian Space Agency provided the OSIRIS-REx Laser Altimeter (OLA), a LIDAR system for topographic mapping.

With both OSIRIS-REx and Hayabusa2 planning to return asteroid samples within a few years of one another, a Memorandum of Understanding (MOU) was signed between NASA and JAXA in November 2014. The agreement included data sharing, the exchange of three mission co-investigators, and a sample exchange between Bennu and Ryugu, enabling the first comparative analysis of two carbonaceous asteroids to offer key insights into the diversity of early Solar System materials.

\subsection{The mission}

OSIRIS-REx arrived at Bennu in December 2018 and began detailed remote-sensing observations. While Hayabusa2 had tracked asteroid Ryugu along its solar orbit, OSIRIS-REx entered gravitationally-bound orbits as close as 1\,km, setting a record for the closest orbit achieved around a small body. 

The selected sampling site was situated in the Hokioi crater, a 20\,m diameter impact feature in the northern hemisphere of asteroid Bennu that appears relatively young. On 20 October 2020, the spacecraft descended to collect a sample and unexpectedly sank nearly half a metre (48.8\,cm) into the surface before the spacecraft thrusters were fired to initiate ascent. The operation left a large 8\,m-wide crater on the asteroid surface.

OSIRIS-REx gathered material via the Touch and Go Sample Acquisition Mechanism (TAGSAM), comprising of a cylindrical sampler head mounted on a 2.8\,m articulated arm (Figure~\ref{fig:orex_spacecraft}). In contrast to the Hayabusa design of capturing ejecta released by a projectile, TAGSAM released a jet of high-purity nitrogen gas for several seconds during downward penetration upon contact with the surface that lifted material into the collection chamber within the TAGSAM head. 24 contact pads on the TAGSAM baseplate also trapped fine particles and dust as the head made contact with the surface. 

\begin{figure*}[!th]
\centering
\includegraphics[width=\textwidth]{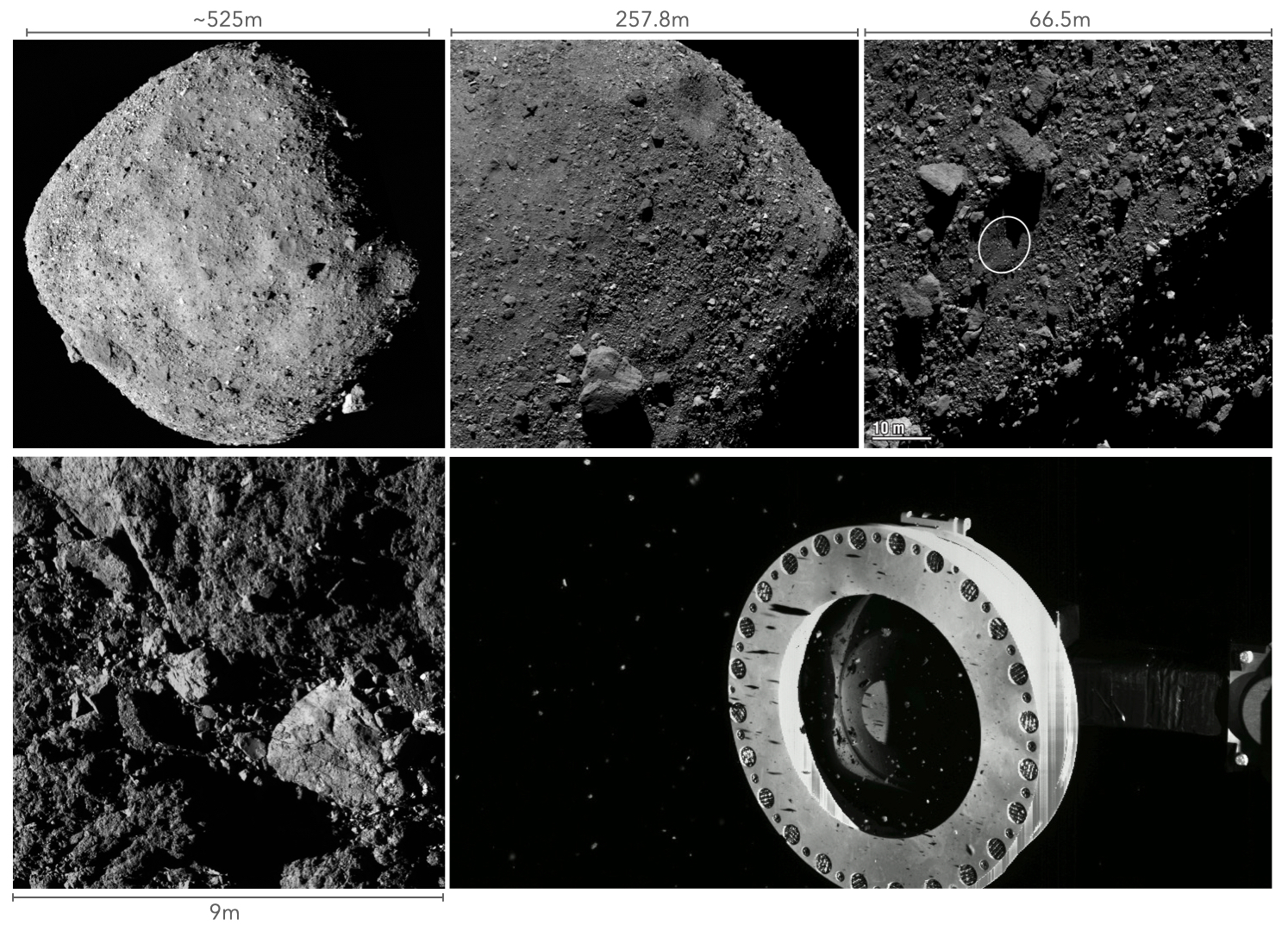}
\caption{Images of asteroid Bennu captured by the OSIRIS-REx spacecraft. Top row (left to right): mosaic of asteroid Bennu composed of 12 PolyCam images, Bennu's southern hemisphere taken with MapCam including the largest boulder, Benben Saxum, Polycam image of the Nightingale sample site. Bottom row: smooth and rough rock types captured with Polycam, TAGSAM after collection images with SamCam, with particles escaping from the sampler head (credit: NASA, Goddard, University of Arizona).} 
\label{fig:bennu}
\end{figure*}

During the imaging of the TAGSAM after the OSIRIS-REx spacecraft ascended, material was seen escaping from the head (Figure~\ref{fig:bennu}). Larger rocks had jammed open the flexible Myler flap that covered the TAGSAM entrance and prevented the expected seal of the main collection chamber. The solution was to quickly stow the TAGSAM head into the sample return capsule, although this action prevented a plan to estimate the quantity of sample by measuring the change in the moment of inertia when the spacecraft was rotated with the TAGSAM on the extended arm empty and then full. 

The OSIRIS-REx reentry capsule landed in the Utah Test and Training Range (UTTR) on 24 September 2023. The descent was tracked with via ground-based and aircraft infrared and optical cameras and the capsule was reached by the recovery team within twenty minutes of landing. The reentry capsule was flown by helicopter to a temporary cleanroom at UTTR where a nitrogen gas purge line was connected, and the following day was transported to the Astromaterials Research and Exploration Science division at the NASA Johnson Space Center in Texas. 

As with the Hayabusa2 mission, contamination control had begun before the spacecraft was launched. Mission requirements included strict limits for non-volatile and organic residue to avoid any compromise of the sample, and materials from the assembly, testing, and launch sites were also archived for comparison during the sample analysis. The OSIRIS-REx cleanroom at the Johnson Space Center conforms to an ISO 5 standard (equivalent to a class 100) and construction materials were selected for low particulate shedding and outgassing properties \citep{Righter2023}. The sample was processed in a clean chamber under pure nitrogen.  

During the disassembly of the TAGSAM head, two of the 35 fasteners could not initially be removed, and about 70\,g of the sample was extracted via the flexible flap. When the remaining bulk sample was recovered, the total sample mass was 121.6\,g, double that of the minimum mission requirement and the largest asteroid sample to be returned to Earth. About 70\% of the Bennu sample has been stored for long-term study. 

After releasing the capsule, OSIRIS-REx moved away from Earth and began an extended mission as OSIRIS-APEX to visit asteroid (99942) Apophis.

\subsection{Asteroid Bennu}

Like Itokawa and Ryugu, (101955) Bennu is a near-Earth asteroid with an orbit between 0.9\,au and 1.36\,au from the Sun. Bennu was discovered by the LINEAR program in September 1999, and received the provisional designation 1999 RQ$_{36}$. The name `Bennu' was selected through a public contest and refers to an ancient Egyptian deity associated with the sun, creation, and rebirth, symbolic of the mission goals to investigate the origins of life. The heron-like depiction of the deity also evokes the shape of the OSIRIS-REx spacecraft with its articulated arm extended for sample collection. The name was formally approved by the International Astronomical Union in 2013.

Classified as a potentially hazardous asteroid, Bennu was extensively studied through visible, infrared, and radar observations before the arrival of OSIRIS-REx. The asteroid was known to have a diameter of about 500\,m with a roughly spheroidal shape and prominent equatorial ridge indicative of a `spinning top' morphology. The surface was expected to be relatively smooth, with a fine-grained ($< 1$\,cm) regolith layer, and only a single identified boulder of size $10-20$\,m \citep{Lauretta2019}.

\subsubsection{A third boulder-strewn rubble pile}
\label{sec:orex_rubble}

Despite being one of the best characterised near-Earth objects, the surface of asteroid Bennu was still highly unexpected. Rather than the predicted smooth surface, the asteroid is dominated by boulders larger than 1\,m, including over 200 boulders exceeding 10\,m. Only small regions less than about 20\,m across appear to be boulder free and covered with fine regolith. Compared to asteroids Itokawa and Ryugu, the cumulative size-frequency distribution of the boulders revealed that Bennu has less surface mass in small particles than Itokawa, while Ryugu maintains the highest number density of boulders $\ge 20$\,m with approximately twice that of Itokawa and Bennu \citep{DellaGiustina2019}. The sole boulder visible in the radar imagery is the most prominent on the Bennu surface. Dubbed `Benben', this boulder stands at 21.7\,m in height and 58\,m wide in the asteroid southern hemisphere (Figure~\ref{fig:bennu}). However, the plethora of smaller boulders were indistinguishable from radar noise. 

Pre-encounter expectations of a fine-grained regolith surface were further based on measurements of the asteroid thermal emission. The data suggested a low bulk thermal inertia, which is typically interpreted as the presence of regolith that would change temperature more rapidly than larger and denser material. OSIRIS-REx confirmed this low value with the OTES and OVIRS spectrometers, which measured a global average of about 350\,tiu during approach. The discrepancy between this low measurement and the boulder-strewn landscape was initially thought to be due to a thin layer of fine particles covering the boulders. However, later proximity measurements of the temperature changes over the asteroid diurnal cycle were more consistent with a nearly dust-free surface dominated by porous boulders with low thermal inertia, similar to Ryugu \citep{Rozitis2020}. 

The contrast between the expectation and reality of the Bennu surface suggests that other carbonaceous asteroids may have significantly rougher structures than previously inferred from ground-based and remote space-based observations. The degeneracy when interpreting even quite detailed ground-based data underscores the critical role of spacecraft reconnaissance and sample return.

Bennu does exhibit the expected spinning-top shape, although with an equatorial ridge that is less prominent than radar imagery had suggested. Measurements with the OSIRIS-REx OLA LIDAR found an equatorial radius of about 270\,m, with a polar-to-equatorial axis ratio of 0.84, giving Bennu a similar shape to asteroid Ryugu but at roughly half the size \citep{Daly2020}. The rotation period of the asteroid is 4.3 hours. Similar to Ryugu (section~\ref{sec:haya2_spinningtop}), this rotation is not swift enough to have produced the equatorial bulge and implies that Bennu once rotated more rapidly, at about 3.5 hours \citep{Sheeres2020}. 

Shape measurements from OLA determined the total asteroid volume to be $6.13 \times 10^7$\,m$^3$. The asteroid mass was derived from tracking small trajectory deflections of the OSIRIS-REx spacecraft due to the asteroid's gravity during slow flybys operations, yielding a mass of $7.3\times 10^{10}$\,kg. Combined, these values gave Bennu an average bulk density of 1.19\,\gcm \citep{Scheeres2019}. Bennu therefore has a volume about 3.5 times that of Itokawa and roughly 1/6 of that of Ryugu, with a density consistent with Ryugu. Analysis of the delivered sample later found an average grain density similar to Ryugu at 1.79\,\gcm. The much lower bulk density of the asteroid compared to the grain density supported what became dramatically apparent during sample collection: asteroid Bennu had a macroporosity of about 50\%, and was therefore also highly porous and a low cohesive rubble pile.

Like Ryugu, the biggest boulders on Bennu are far too big to be ejecta from the largest craters, with sizes exceeding 1/10 of the asteroid diameter. This suggests that these boulders are original fragments of a parent body that disintegrated before re-accumulating into the rubble pile now known as Bennu. Less like Ryugu is a striking albedo dichotomy among the Bennu boulders, which points to two distinct lithologies. Larger boulders have albedos close to the asteroid's very dark global average of about 4.4\,\% at 0.55\,\micron. In contrast, many smaller boulders with diameters less than 10\,m are much brighter, producing a global albedo range stretching over a factor of four from 3.3\,\% to 15\,\% \citep{DellaGiustina2019}. This is a significantly greater variation than the surface of Ryugu or Itokawa. Thermal inertia measurements reinforce this distinction, with the high-reflectance boulders exhibiting values of 400 - 700\,tiu, while regions populated with the low-reflectance boulders show 180 - 220\,tiu \citep{Rozitis2020}. This suggests that the brighter boulders could be formed from less porous material, consistent with more angular surface compared to the hummocky appearance of their darker counterparts. Such an interpretation was confirmed with the delivered sample, which contains both hummocky and angular grains thought to originate from the two boulder morphologies. The angular grains are denser and may have formed under higher-pressures deeper within the parent asteroid (Lauretta \& Connolly et~al.~\citeyear{LaurettaConnolly2024}, Connolly \& Lauretta et~al.~\citeyear{ConnollyLauretta2025}).

While Ryugu does not display the same marked albedo variations, a main population of porous low density boulders and a minor group of angular higher density boulders are observed, and are likewise suspected of forming at different pressures within the asteroid. As noted in section~\ref{sec:haya2_rubblepile}, highly porous material would likely not survive atmospheric reentry to Earth. Thus, even the rarest carbonaceous chondrite meteorites probably fail to reflect the true bulk composition of carbonaceous asteroids. The delivered samples are therefore valuable examples of the typical composition, including both more fragile material and material samples from different depths in the parent asteroid. 

The structures of Itokawa, Ryugu, and Bennu support the hypothesis that Near-Earth Objects between a few hundred metres to a few kilometres in size may typically be rubble piles, formed by the disruption and loose accumulation of material from larger parent bodies that originated elsewhere in the Solar System. This has major scientific implications, as exploring other primitive NEOs could provide a more complete cross-section of the early planet-forming material from across the Solar System. It is also critical for planetary defence, as any attempt to deflect a hazardous NEO will require precise knowledge of the internal structure.

\subsubsection{An active asteroid}
\label{sec:orex_particles}

In early 2019, an unexpected discovery was made by the navigation cameras onboard OSIRIS-REx. Asteroid Bennu was periodically ejecting particles with diameters $<1 \sim 10$\,cm \citep{Lauretta2019}. This phenomenon had previously been observed in `active asteroids': small bodies with asteroidal orbits in the asteroid belt or near-Earth space that exhibit transient or sustained emission of dust or gas. Although Bennu shares spectral similarities with known active asteroids, no activity had been seen in ground-based observations or during the spacecraft approach. The lack of detection was due to the small size of the emitted particles. Apart from the occasional larger fragment, the emission from Bennu was below instrument sensitivity except at close range. 

During the observation campaign in 2019 when OSIRIS-REx was about 1.5 - 2.0\,km from Bennu's centre, three major ejection events each releasing about $70 \sim 200$ particles were recorded, along with several smaller events ($<20$ particles). A persistent particle background around Bennu was also detected, likely from these and past ejections. 

The three main events originated from different regions on Bennu with no obvious geological significance. Many particles lacked sufficient velocity to escape the gravity of Bennu and reimpacted the asteroid surface, while faster particles escaping entirely. On comets, dust release occurs via the gas drag during the ice sublimation. A similar process was considered for Bennu, but this would require the presence of ice close to the asteroid surface. Several of the observed ejection events occurred at low latitudes where temperatures would be too high ($\sim 390$\,K) for such ice to be stable. A number of alternative mechanisms for Bennu have been proposed. Instead of ice, volatile release could be occurring from the thermal dehydration of hydrated minerals. Mechanical stresses such as grinding can release trapped water molecules and the structural -OH in the phyllosilicates, which sublimate to provide the gas pressure to release dust. Micrometeorites are another possible source, with low-collision impacts that avoid crater or plasma development, but impart enough energy to release dust. A third option is dust released through thermal stress. The temperature gradient down through the top surface layers of Bennu could drive cracks in the rocks which eventually disintegrate and eject particles. 

Dust production from carbonaceous active asteroids has been hypothesised as another delivery mechanism for organics to the early Earth. Evidence includes the active asteroid (3200) Phaethon, whose annual Earth approach is thought to produce the Geminids meteor shower. Estimating the frequency of active asteroids and the mass shed in dust is a key constraint on the dust-delivery theory. Bennu's activity suggests that active asteroids may be more common and more productive than previously believed, with smaller ejection events likely missed from Earth. Phaethon will be the target of a fly-by observation with the JAXA DESTINY+ mission which is currently scheduled to launch in 2028. Bennu itself could eventually generate a similar meteor shower. While no such shower has been yet detected, simulations suggest that ejecta from Bennu would take about 450 years to evolve into an annual meteor display. Such a shower may therefore begin in the future, depending on when Bennu became active \citep{Ye2019}. 

\subsubsection{The future orbit of asteroid Bennu}

Prior to the arrival of the OSIRIS-REx spacecraft, the trajectory of asteroid Bennu was predicted from extensive ground-based tracking data. The dominant non-gravitational force on the asteroid orbit was identified as the thermal radiation recoil from the Yarkovsky effect \citep{Chesley2014}. Accurate orbit estimation is particularly important, as Bennu will make a close approach to Earth in the year 2135. Should the asteroid pass through small `gravitational keyholes' during this close approach, the gravity of the Earth could deflect the asteroid onto a future impact trajectory. As keyholes are tiny, risk assessment requires extremely precise modelling that
includes drift from factors such as the Yarkovsky effect.

Remote-sensing data from OSIRIS-REx greatly improved the thermophysical model for Bennu by providing accurate measurements of the asteroid shape, spin rate, albedo, thermal inertia, and surface roughness. This yielded a refined Yarkovsky acceleration, which was validated by metre-level constraints on the position of the asteroid from the spacecraft tracking data \citep{Farnocchia2021}. The updated value corresponds to a semimajor axis drift of -284.6\,m yr$^{-1}$, which will vary as the orbit of Bennu evolves through to 2135. The refinement constrained the cumulative impact probability through the year 2300 to about 1 in 1750 (or 0.057\%). 

The Yarkovsky model for Bennu also serves as a reference to predict the radiative forces on asteroids not visited by spacecraft by leveraging scaling relationships, as the Yarkovsky effect is inversely proportional to the asteroid diameter \citep{Hung2023}. While this assumes similar properties such as density and albedo, Bennu's well-characterised parameters provide more reliable limits. The delivered sample will further tighten these constraints through laboratory measurements for the thermal and physical properties, improving interpretation of remote data for other asteroids. 

\subsubsection{A B-type asteroid and a carbonaceous chondrite}

During the approach to asteroid Bennu, the OSIRIS-REx spacecraft replicated the test performed by Hayabusa2: the onboard OVIRS spectrometer measured a disc-integrated spectrum to compare with ground-based observations. The two datasets closely agreed, classifying asteroid Bennu as a B-type carbonaceous asteroid based on its blue-sloped spectrum \citep{Hamilton2019}. 

At longer wavelengths than those accessible from the ground, OVIRS detected a near-infrared absorption feature centred at 2.74\,\micron. This feature appeared both in the disc-integrated and spatially resolved spectra, indicating wide-spread hydrated minerals containing the hydroxl (-OH) groups. From this, it was thought that Bennu resembles CI, CM, or CR chondrite meteorites, with an early assessment based on the shape of the absorption feature in the remote analysis favouring strongly aqueously altered CM chondrite types \citep{Hamilton2019}. 

\begin{figure*}[!th]
\centering
    \includegraphics[width=\textwidth]{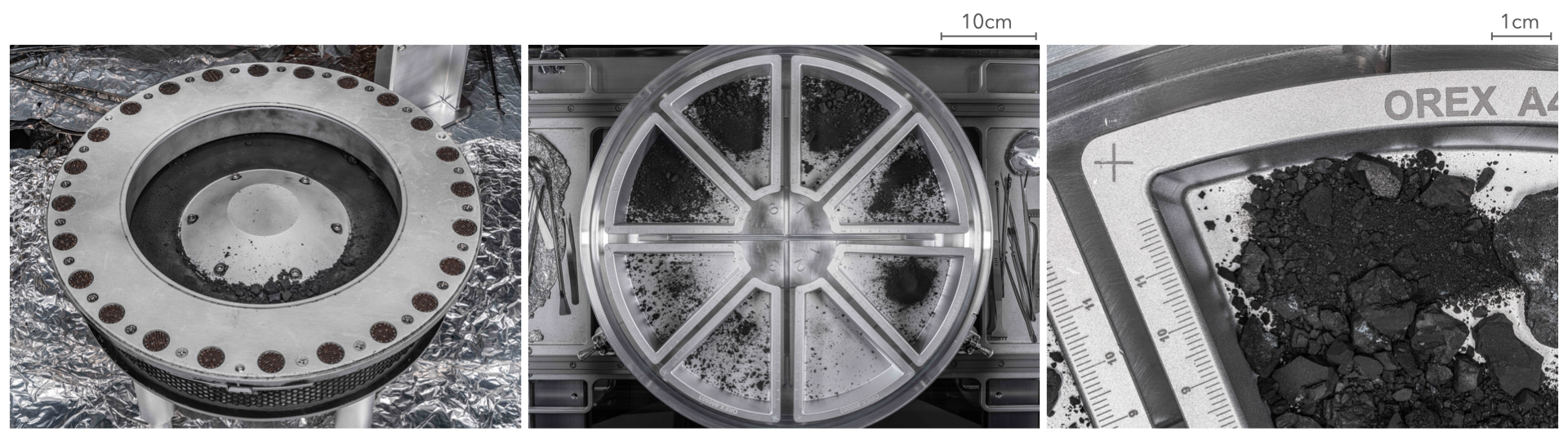}
\caption{The sample delivered from asteroid Bennu. Left to right: The TAGSAM sampling head and contact pads, eight sample trays containing 51.2g of material from asteroid Bennu, close up view of the top left corner of one of the deep sample trays (credit: NASA AIVA/Blumenfeld/Aebersold).} 
\label{fig:bennu_sample}
\end{figure*}

Analysis of the delivered sample in 2023 (Figure~\ref{fig:bennu_sample}) confirmed extensive aqueous alteration abundant hydrated minerals. However, detailed laboratory measurements revealed that the Bennu sample most closely aligned with CI chondrites. Measured abundances of 54 elements from lithium to samarium showed an average chemical composition with higher abundances of volatiles than CM or other carbonaceous meteorite types, and with a refractory element concentrations within 5\% percent of solar values (Barnes \& Nuygen et~al.~\citeyear{Barnes2025}). The composition was very similar to that of asteroid Ryugu, although the Ryugu sample exhibited a slightly greater excess of refractory elements compared to Bennu or CI chondrites (Figure~\ref{fig:ryugu_elements}) (Lauretta \& Connolly et~al.~\citeyear{LaurettaConnolly2024}). 

The oxygen isotope analysis of Bennu also resembled the Ryugu sample, with an average elevation of $^{17}$O compared to other carbonaceous meteorite groups, and an absence of both sulphates and ferrihydrite (section~\ref{sec:haya2_CIchondrite}). This strengthens the suggestion that these features in CI chondrites are due to contamination from the terrestrial environment. 

Although both asteroid Ryugu and Bennu were chosen for their suspected carbonaceous nature, the close association of both asteroids with rarest carbonaceous meteorite class initially seems surprising. Remote sensing and sample analysis from both asteroids now suggest that the chemically pristine CI chondrite material is common is space but rarely survives atmospheric entry, explaining its scarcity among meteorites found on Earth. 

\subsubsection{Minerals: evidence of ancient brines}
\label{sec:orex_aqueous}

While the remote-sensing data from Hayabusa2 had initially suggested that asteroid Ryugu had a dehydrated carbonaceous composition, observations by OSIRIS-REx indicated that the surface of Bennu is abundant in hydrated minerals that appeared to have undergone minimal heating \citep{Hamilton2019}. Based on these observations, it was expected that the Bennu sample would be dominated by phyllosilicates.

The sample returned from Ryugu would prove to contain significantly more hydrated material than spacecraft observations had suggested (section~\ref{sec:haya2_aqueous}). In contrast, the Bennu sample strongly aligns with the predictions from the remote-sensing data. Consistent with the CI chondrite composition inferred from the chemical element abundances, approximately 80\,\% of the Bennu sample volume consists of hydrated phyllosilicates, primarily the magnesium-bearing serpentine and saponite. Interspersed with the phyllosilicate matrix are smaller quantities of minerals that precipitate from aqueous solution, including magnetite, sulphides such as pyrrhotite, and carbonates such as dolomite, calcite, and iron-rich magnesite (see Table~\ref{table:bennu_minerals}). Both the mineral types and abundances are similar to those found in asteroid Ryugu, and are a close match to CI chondrite meteorites (Lauretta \& Connolly et~al.~\citeyear{LaurettaConnolly2024}). 

\begin{table*}[t]
\caption{Abundance (by volume percentage, vol\%) of the main minerals in a Bennu
aggregate sample (Lauretta \& Connolly et~al.~\citeyear{LaurettaConnolly2024}).}
\label{table:bennu_minerals}

\vspace{4pt}

\centering
\begin{tabular*}{\textwidth}{@{\extracolsep{\fill}} l l r}
\hline
Mineral group & Formula & $\sim$vol\% \\
\hline
Phyllosilicates &  & 80 \\
\hspace{3mm}Serpentine & (Mg,Fe)$_3$Si$_2$O$_5$(OH)$_4$ &  \\
\hspace{3mm}Saponite & Ca$_{0.25}$(Mg,Fe)$_3$((Si,Al)$_4$O$_{10}$)(OH)$_{2n}$(H$_2$O) &  \\[1ex]

Fe sulphides &  & 10 \\
\hspace{3mm}Pyrrhotite & (Fe,Ni)$_{1-x}$S &  \\
\hspace{3mm}Pentlandite & (Fe,Ni)$_9$S$_8$ &  \\[1ex]

Magnetite & Fe$_3$O$_4$ & 5 \\[1ex]

Carbonates &  & 3 \\
\hspace{3mm}Dolomite & CaMg(CO$_3$)$_2$ &  \\
\hspace{3mm}Magnesite & (Mg,Fe,Mn)CO$_3$ &  \\
\hspace{3mm}Calcite & CaCO$_3$ &  \\[1ex]

Anhydrous silicates &  & 2 \\
\hspace{3mm}Olivine & (Mg,Fe)$_2$SiO$_4$ &  \\
\hspace{3mm}Pyroxene & (Mg,Fe)SiO$_3$ &  \\[1ex]

Phosphates &  & trace \\
\hline
\end{tabular*}
\end{table*}

With Bennu too small to sustain liquid water, aqueous alteration must have occurred on the parent body that accreted from rock and ice in the outer Solar System. Temperatures during the aqueous production of these minerals were estimated from the Bennu sample by comparing two iron-nickel sulphides: pyrrhotite ((Fe,Ni)$_7$S$_8$) and pentlandite ((Fe,Ni)$_9$S$_8$). The composition of these sulphides are sensitive to the formation temperature, which can be inferred from the position of tie lines connecting these minerals plotted on an iron-nickel-sulphur (Fe-Ni-S) phase diagram. The composition in the Bennu sample suggests alteration at approximately 25\degC, with an upper limit near 100\degC (Zega \& McCoy et~al.~\citeyear{Zega2025}). 

However, not all minerals in the Bennu sample have experienced aqueous alteration. A few percent of the sample volume consists of anhydrous silicates such as forsteritic olivine (Mg$_2$SiO$_4$) and low-calcium pyroxene ((Mg,Fe)SiO$_3$), remnants of the original unaltered material accreted by the asteroid. This proportion is higher than that found in the Ryugu sample, indicating that Bennu's parent body was not fully hydrated into secondary minerals (Lauretta \& Connolly et~al.~\citeyear{LaurettaConnolly2024}, Barnes \& Nuygen et~al.~\citeyear{Barnes2025}). Further evidence is found in the hydrogen isotope composition of organic molecules in the Bennu sample, which show a relative enrichment of deuterium compared to the Ryugu sample or other CI and CM chondrites. Studies of carbonaceous chondrites have shown that the deuterium to hydrogen ratio (D/H) in insoluble organics decreases with increasing aqueous alteration. The elevated deuterium abundance in the Bennu sample therefore supports the conclusion that the parent asteroid was not completely hydrated (Barnes \& Nuygen et~al.~\citeyear{Barnes2025}). 

Despite broad agreement with the remote sensing data, the Bennu sample was not free of surprises. An early unexpected find was the presence of magnesium-sodium phosphates that had not been observed by the spacecraft. Small quantities of similar phosphates were previously found in the sample from asteroid Ryugu and in certain meteorites, but those in the Bennu sample were notable for their relatively large size  and absence of inclusions from other minerals (Lauretta \& Connolly et~al.~\citeyear{LaurettaConnolly2024}). Unlike the water-rock reactions that produced the hydrated minerals and precipitated the magnetite and sulphides, phosphates are evaporite minerals that are commonly classified as salts. Salts are highly soluble in aqueous or water-rich fluids and typically remain in solution until concentrations rise, usually as the fluid begins to evaporate. The abundance of phosphate minerals in the Bennu sample therefore not only confirms the presence of water on the parent asteroid, but also suggests a prolonged period of geochemical activity that enriched the water in phosphate. It is also the first evidence that Bennu once hosted brines. 

Brines are liquid water with a high concentration ($> 3.5$\,\% by weight) of dissolved solids. They form through evaporation or freezing, processes that reduce water content without removing the dissolved compounds. This concentration process promotes the formation of more complex molecules and lowers the freezing point to maintain liquid water in cold environments. The enriched solution also facilities the precipitation of highly soluble compounds such as salts, whose presence is a strong indicator of past brine activity and is of particular interest for astrobiological studies. However, studying brine formation via meteorites is challenging because any salts rapidly absorb moisture from the Earth's atmosphere and dissolve. Salts that have been found in meteorites such as Winchcombe therefore result from terrestrial contamination. 

While the Ryugu sample contained fewer phosphates, other salts such as sodium carbonate, sodium chloride, and sodium sulphate were identified \citep{Matsumoto2024}. These too are thought to have formed from brines as Ryugu's parent body lost its water through evaporation or freezing. Which process ended Ryugu's water-rich past remains unclear, but Bennu offers more clues. Further analysis of the Bennu sample revealed not only phosphates but also a diverse inventory of salts that included sodium-rich carbonates, sulphates, chlorides and fluorides, and potassium chloride (McCoy \& Russell et~al.~\citeyear{McCoy2025}). Together, these minerals are evidence of an evaporite sequence in which salts progressively precipitated as water slowly disappeared from the system. Comparison with similar evaporite sequences in terrestrial brines suggests that temperatures of about $20 -29$\,\degC could have driven this process through evaporation rather than freezing, which is consistent with the temperature conditions inferred from Bennu's sulphide mineralogy. The Bennu parent body therefore probably lost its liquid water and ice before migrating into the inner Solar System, carrying instead the hydrated minerals, salts, and compounds produced during the aqueous alteration and the subsequent evaporation. This evolutionary path may help resolve a long-standing question about water distribution in the inner Solar System: the inward migration of ice-rich bodies could deliver too much water to the region, but this risk is mitigated if the water is reduced or lost through evaporation \citep{Sekine2025}.

Evidence for the formation location of the Bennu parent was found in high concentrations of ammonia in the sample, which is 75 times higher than for the Ryugu sample and greater than most carbonaceous chondrites (Glavin \& Dworkin et~al.~\citeyear{Glavin2025}). This suggests that the parent body originated past Jupiter's current orbit in the outer Solar System, where ammonia ice was stable. Radiogenic heat would have melted the ice and triggered extensive aqueous alteration. As the water began to evaporate through fractures in the rock, the salt-rich brines would have formed. The abundance of ammonia could have allowed some of these brines to persist even at low temperatures after the exhaustion of the radiogenic elements.

Remote sensing data suggests that similar salt minerals to those found on Bennu and Ryugu may exist on the dwarf planet Ceres and the Saturn moon Enceladus, both of which are thought to harbour subsurface brines. Ammonium salts have also been identified on Ceres, suggesting a possibly similar formation history to Bennu. The evolution of such icy worlds is of significant biological interest, as it is speculated that such environments could synthesise the main ingredients for life. Bennu may represent an early stage in this process, and the evidence of ancient brines in carbonaceous asteroids offers an exceedingly rare opportunity to study the organic synthesis that could occur in these environments. 

\subsubsection{Organics}

The aqueous alteration of the Bennu mineralogy and parallels with icy worlds hinted at a diverse organic inventory. Early examination of the returned sample measured a bulk carbon content of $4.54$\% by weight, with an estimated 90\% in organic molecules, and a nitrogen content of 0.23 - 0.25\% by weight. These abundances are higher than those found in the samples from Ryugu or previously studied carbonaceous meteorites (Lauretta \& Connolly et~al.~\citeyear{LaurettaConnolly2024}).

Analysis of the organic compounds confirmed the early expectations. 16,000 molecular formulae were identified that consisted of carbon, hydrogen, nitrogen, oxygen, sulphur and magnesium. A total of 33 amino acids were detected, including 14 out of the 20 standard proteinogenic amino acids that are used in terrestrial biology, with glycine being the most abundant. All five primarily nucleobases used in DNA and RNA (adenine, guanine, cytosine, thymine, and uracil) are also present, alongside 18 other N-heterocycles that share the nitrogen-containing ring structure (Glavin \& Dworkin et~al.~\citeyear{Glavin2025}). 

The diversity of organics is similar to that in the Ryugu samples, but with a more nitrogen-rich rather than sulphur-rich composition. This likely reflects the abundance of ammonia, which would have facilitated the production of amino acids and N-heterocycles for which ammonia is considered a potential precursor for synthesis. A nitrogen-rich composition could indicate a period of low-temperature aqueous conditions where the highly volatile nitrogen compounds could be retained. After short-lived radionuclides that provided early internal heating were exhausted, high concentrations of dissolved ammonium salts could have maintained liquid brines even at temperatures as low as 176\,K. This evolutionary pathway may resemble that of icy worlds such as Ceres, suggesting the presence of similar organics on the dwarf world. 

Comparison of amines (ammonia derivates where one or more hydrogen atoms has been a replaced by an organic group) revealed more equal abundances of less stable forms in the Bennu sample compared to Ryugu. Trends in meteorites suggest this indicates a lower degree of aqueous alteration, as extensive alteration can destroy or transform amines into other compounds and reduce their abundance and diversity. The preservation of a wider range of amines in Bennu therefore supports the interpretation of a less aqueously altered parent body compared to Ryugu, consistent with the higher abundance of anhydrous minerals and deuterium enrichments (section~\ref{sec:orex_aqueous}). This may also have allowed a larger distribution and abundance of amino acids (Barnes \& Nuygen et~al.~\citeyear{Barnes2025}). 

The chiral amino acids in the Bennu sample were present as a racemic mixture, with equal abundances of D- and L- enantiomers. This agrees with findings from the Ryugu sample, which also contained equal quantities of both chiral forms (section~\ref{sec:haya2_organics}). The results challenge the hypothesis that the preference of life to favour L-enantiomer amino acids originated in a bias in extraterrestrial materials. The reason for the left-handed bias therefore remains unresolved. 

\subsubsection{Presolar material}

The close solar composition and indirect evidence of heating from short-lived radionuclides suggests that like Ryugu, Bennu's parent body was among the earliest to form in the Solar System. Although the mineralogy reflects a water-rich evolutionary history, remnants of material predating the Sun were found in the returned sample. Anomalous oxygen, carbon, and nitrogen isotopic compositions were found in submicrometre-sized grains embedded in the phyllosilicate matrix, while similar outliers in hydrogen, nitrogen, and carbon isotopes were detected in organic matter. These anomalous signatures closely resemble those in the Ryugu sample, and indicate a presolar origin (section~\ref{sec:haya2_presolar}). The presolar grains likely formed in evolved stars and supernovae, while the small quantities of isotopically unusual organics may have originated in the molecular cloud or outer regions of the protoplanetary disc (Barnes \& Nuygen et~al.~\citeyear{Barnes2025}). Because presolar grains are easily destroyed by heat and hydration, their survival in the Bennu and Ryugu samples supports the assertion that neither asteroid experienced high temperatures, or complete aqueous alteration. 

Not all isotopic anomalies are attributed to presolar material. Submillimetre anhydrous silicates in the Bennu sample exhibit both $^{16}$O-rich and $^{16}$O-poor compositions, respectively similar to amoeboid olivine aggregates (AOAs) and chondrules. These minerals are commonly found as large refractory inclusions in most types of carbonaceous chondrites, but are rare in CI chondrites and in the samples from Ryugu and Bennu. AOAs and chondrules are high-temperature condensates that are thought to have formed very early in the hotter inner Solar System. The presence of only small fragments in the delivered samples suggests that the Bennu and Ryugu parent bodies accreted in regions where these inclusions were scarce, supporting a formation location in the colder outer Solar System (Barnes \& Nuygen et~al.~\citeyear{Barnes2025}). 

\subsubsection{Space weathering}

Although Ryugu and Bennu share similar chemical and mineralogical compositions, the reflected spectra show opposite slopes: Ryugu is red-sloped, while Bennu is blue-sloped. Even more intriguingly, smaller and younger craters on Bennu that display more recently exposed material appear redder than the global average. This suggests that space weathering on Bennu is creating a bluer surface, while the same process on Ryugu is causing reddening. 

This hypothesis was tested by the surface disruption caused by the OSIRIS-REx sample collection. The Hokioi crater revealed a spectrally redder region after the spacecraft touchdown than the surroundings. The delivered sample, collected from both surface and sub-surface as OSIRIS-REx sank into the asteroid, display an even stronger red-sloped spectrum.  

Differences in space weathering processes have traditionally been linked to variations in surface composition. On the anhydrous Moon and Itokawa, micrometeorite impacts vaporise grain surfaces to form a smooth layer of redeposited material. In contrast, smooth layers on the Ryugu phyllosilicate grains are produced primarily by solar wind irradiation (section~\ref{sec:haya2_spaceweathering}). Yet both Ryugu and Bennu are carbonaceous asteroids with phyllosilicate-dominated surfaces, making the contrasting weathering trends puzzling.

Clues emerged from estimates of the asteroid surface exposure age, which represents the time that material remains in the uppermost layer. Even without a major impact, surface material on asteroids moves and slides to create a turnover of the layer exposed to space weathering. The density of microcraters and ionisation tracks caused by energetic solar wind particles provide a measure of how long the sample grains have been exposed. The results indicated that the surface of Bennu has been exposed to space weathering an order of magnitude longer than that of Ryugu (Keller \& Thompson et~al.~\citeyear{Keller2025}). This suggests that carbonaceous asteroids may follow a weathering sequence that initially causes reddening, followed by the spectrum becoming bluer with prolonged exposure. This differs from anhydrous material, where weathering continues to redden the surface. Understanding these different trends could guide estimates of age for asteroids in the main belt. 

\subsubsection{A common parent?}

As two rubble-pile NEOs believed to have formed from the catastrophic disruption of aqueously altered parent bodies, could Ryugu and Bennu be fragments of a common progenitor asteroid? Dynamical models suggest that Bennu once orbited in the inner main belt, where the new Polana asteroid family provides the closest spectral match. Ryugu exhibits a similar dynamical link, suggesting that both asteroids could be fragments of a single progenitor that disrupted to create the new Polana family \citep{Arredondo2025}. 

Countering this hypothesis is the significant compositional differences in the delivered samples. Bennu contains abundant ammonia and nitrogen-bearing species, whereas Ryugu exhibits carbonate fluid inclusions and signatures of more extensive aqueous alteration. This is suggestive of two parent bodies, with Bennu's parent originating in colder, more distant regions of the solar nebula, where highly volatile nitrogen compounds remained stable. 

However, these differences could be reconciled by a single parent body with radial thermal and compositional gradient. Ryugu may have formed from material in the warmer interior, where aqueous alteration was more complete and nitrogen volatiles loss occurred. Bennu could originate from fragments of the cooler outer layers that preserved nitrogen-rich phases \citep{Noguchi2025}. Subsequent collisional evolution and varying residence times in near-Earth space would have modified the surface of each asteroid by differing degrees of space weathering.

\subsubsection{History of Bennu: the possible tale of two asteroids}

\begin{figure*}[!th]
\centering
\includegraphics[width=\textwidth]{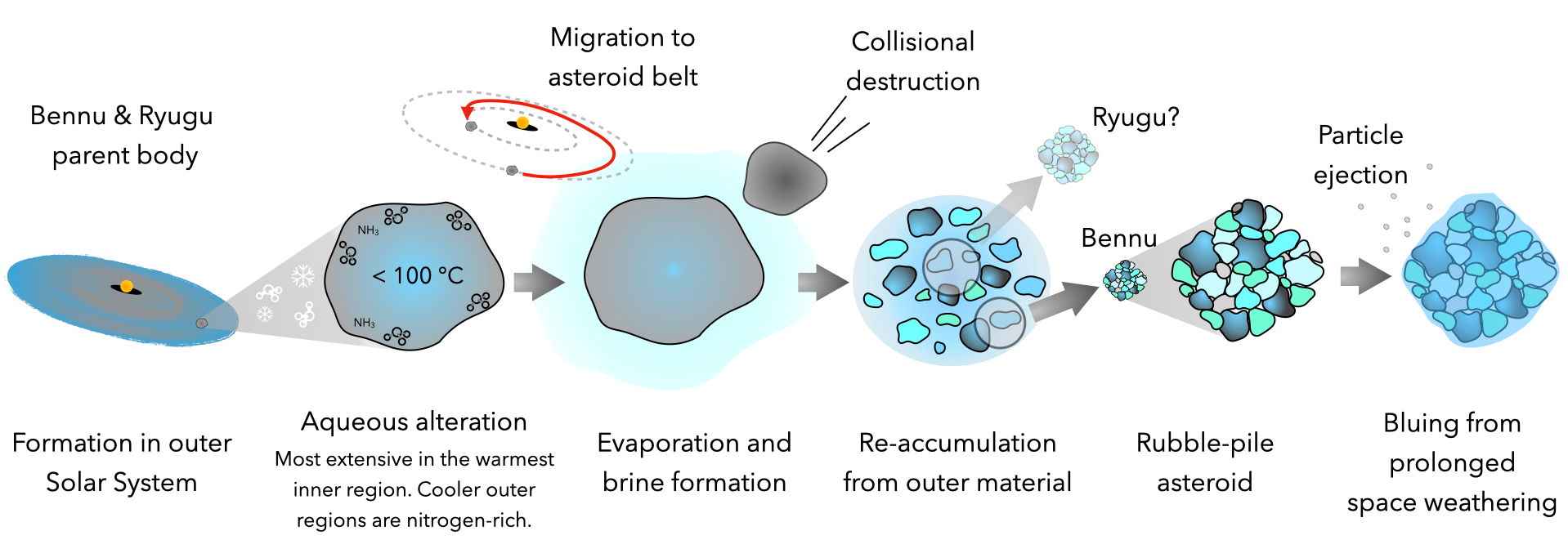}
\caption{The history of asteroid Bennu.} 
\label{fig:bennu_history}
\end{figure*}

Like Ryugu, the story of asteroid Bennu traces back to some of the earliest solid material that formed around the young Sun. The parent body of Bennu was a planetesimal that probably originated in the outer Solar System, beyond the current orbit of Saturn. In this region, temperatures would have been low enough to allow the accretion of ammonia-rich ices, along with the incorporation of presolar grains and organic compounds. It is possible that this planetesimal was the same carbonaceous body that would eventually also produce asteroid Ryugu, or it may have been a separate example of the primitive material that would form the planets.  

As the interior of the planetesimal warmed due to the decay of short-lived radionuclides such as $^{26}$Al, a water-rich fluid formed and drove extensive aqueous alteration. This process transformed the mineralogy into a phyllosilicate-rich matrix containing sulphides, magnetite, and carbonates, and a racemic mix of organic molecules formed. Highly volatile nitrogen compounds may have been lost from the hottest central regions, while the cooler outer layers might have remained less aqueously altered and hosted an even more diverse set of nitrogen-rich organics. 

Over time, the parent body migrated inward toward the asteroid belt. Water gradually evaporated, forming brines that could remain liquid due to low freezing points even as the radionuclide heat began to deplete. Salts precipitated from these concentrated solutions in a chemical system that might be a precursor for icy worlds such as Ceres or Enceladus. 

Eventually, a major collision shattered the planetesimal, creating a group of smaller bodies that possibly became the new Polana asteroid family. Fragments coalesced into rubble pile asteroids, which underwent successive generations of collision disruption and reassembly. Bennu  formed from material where the nitrogen rich chemistry had flourished. If Ryugu originated from the same parent body, then it would have coalesced from material deeper within the planetesimal. Subsequent Yarkovsky and YORP effects evolved the asteroid into a spinning-top shape and nudged Bennu into a near-Earth orbit. Bennu may have reached the NEO orbit before Ryugu, and developed a bluer spectrum as the top layer of regolith experienced prolonged space weathering. 

Today, Bennu and Ryugu preserve a record of the water-altered minerals and organics in early Solar System material and a transportation mechanism from the outer Solar System to the terrestrial planet region.

\section{Conclusions}

Three pioneering missions have successfully returned asteroid samples to Earth--a feat that remains one of the most demanding in space exploration. Each mission faced a unique set of obstacles, requiring ingenuity and an adaptable design to succeed. The comparison between the spacecraft remote-sensing data and laboratory analyses of the delivered samples provided the first unequivocal data points connecting meteorite studies to ground-based observations. The findings further revealed a detailed narrative about some of the earliest solid materials in our Solar System: the primordial building blocks out of which the planets formed. Key insights from the stories of Itokawa, Ryugu, and Bennu include,

\begin{itemize}
  \item \textbf{Rubble-pile structure:} Asteroids Itokawa, Ryugu, and Bennu all have a rubble pile structure consisting of a loosely aggregated material. This supports the hypothesis that Near-Earth Objects (NEOs) between a few hundred metres to a few kilometres in size are typically rubble-piles. Carbonaceous NEOs may be particularly porous, with low-density boulders. This is important for interpreting ground-based observations and for planetary defence strategies. 
  \item \textbf{Primordial time capsules:} The rubble-pile structure is evidence that the asteroids are fragments of larger parent bodies that were catastrophically disrupted. The compositions of the asteroids preserves a record of the parent, especially in the case of carbonaceous asteroids that have experienced minimal thermal metamorphism, providing a time capsule of primordial planetary material.
  \item \textbf{Meteorite-asteroid link:} S-type asteroids such as Itokawa closely match ordinary chondrite meteorites, while C-type and B-type asteroids resemble carbonaceous chondrites. Notably, Ryugu and Bennu are examples of a common asteroid spectral class, yet the delivered samples are a close match to very rare CI carbonaceous chondrites. This suggests a strong bias in the meteorite population, as the majority of carbonaceous asteroid material is likely not surviving atmospheric entry.
  \item \textbf{Space weathering effects:} Space weathering can alter the asteroid spectra, causing a reddening or bluing of the surface layer which can mask the bulk composition of the asteroid. This must be considered when interpreting ground-based or remote-sensing observations.
  \item \textbf{An early outer Solar System origin:} Ryugu and Bennu may originate from a common parent celestial body that formed just a few million years after the start of the Solar System, likely in the outer region beyond the current orbits of Jupiter and Saturn. The parent body then migrated to the asteroid belt before fragmenting in a major collision. This may have been a major pathway for delivering outer Solar System material to the terrestrial planet region. 
  \item \textbf{Water and organics:} The parent body of Ryugu and Bennu contained water-rich fluids which produced hydrated minerals and a wide diversity of organic molecules, including those used for biological activity on Earth. Delivery of such material to the early Earth may have contributed to the development of habitability. However, amino acids were present as a racemic mix, indicating that the biological preference for left-handed enantiomers did not originate from a biased extraterrestrial source. 
  \item \textbf{Implications for habitable worlds:} The presence of organic molecules in early Solar System bodies suggests that biological relevant compounds were wide-spread. Further studies of Bennu and Ryugu could cast light on the formation of non-Earth-like habitable environments, such as icy moons. The abundance of organic molecules in early planetesimals also implies that similar compounds may be common in other planetary systems.  
\end{itemize}

\section{The future}

The analysis of the samples from Itokawa, Ryugu, and Bennu continues worldwide. In the coming years, a major focus will be comparative studies between the Ryugu and Bennu samples, which are expected to provide further insights into the history of these asteroids and the diversity of Solar System material. 

Two new small body sample return missions will take place in the next few years. In 2025, China launched Tianwen-2, targetting a sample return from S-type asteroid 469219 Kamo'oalewa in 2027 followed by a rendezvous with the active asteroid 311P/PanSTARRS. The next JAXA sample return mission, the Martian Moons eXploration (MMX) mission, is scheduled for launch in FY2026, aiming to return material from Mars's moon Phobos in 2031 in what will be the first sample return from the Mars system. While strictly not an asteroid mission, the origin of the two Martian moons remains debated, with one leading hypothesis suggesting that the pair are captured D-type asteroids which are thought to be rich in organics. Samples from this region that sits at the gateway between the inner and outer Solar System is expected to reveal insights into Mars's early formation and more about the composition of material transported into the terrestrial planet zone. 

A longer-term project being considered is a sample return from a comet. Comets form in the volatile-rich outer Solar System and have spent most of their history in the Kuiper Belt or Oort Cloud, making these small bodies the least altered building blocks of the Solar System beyond the ice line. Returning cometary material poses significant challenges due to its high volatile content, which must be preserved during the journey back to Earth. Both JAXA (Next Generation small-body Sample Return Mission, NGSR) and NASA (Comet Astrobiology Exploration Sample Return, CAESAR) have proposed concepts for such a mission. Meanwhile, the ESA-led Comet Interceptor mission will launch in 2029, with the aim to rendezvousing with a long-period comet or interstellar object for remote analysis.

The Hayabusa2$\sharp$ and OSIRIS-APEX spacecraft are now on extended missions with a planetary defence focus. Hayabusa2$\sharp$ will flyby NEO (98943) Torifune (2001 CC21) in July 2026 to observe this S-type asteroid and test technology relevant to planetary defence manoeuvrers. The spacecraft then plans to rendezvous in 2031 with 1998 KY26, a small rapidly rotating asteroid, which is a type not previously investigated by spacecraft. 

OSIRIS-APEX will rendezvous with asteroid Apophis shortly after the asteroid completes a close approach to Earth in 2029. The approach by an asteroid the size Apophis is an event that only occurs around once in a 1000 years, offering an effectively unique opportunity to study the gravitational effects of a planet on a near-Earth object. The ESA/JAXA Ramses mission will also rendezvous with Apophis during this event, launching in 2028 alongside the JAXA DESTINY+ mission, which will first flyby Apophis and later visit other NEOs, including the active asteroid (3200) Phaethon.

\section*{Acknowledgements}

The authors would like to thank Makoto Yoshikawa for his helpful advice.

\section*{Citation}

\noindent E.~J. Tasker, H.~C. Connolly Jr \& S. Tachibana (08 Apr 2026):
The science from asteroid sample return missions, Contemporary Physics, DOI: \href{https://www.tandfonline.com/doi/full/10.1080/00107514.2026.2646056}{10.1080/00107514.2026.2646056}

\bibliographystyle{abbrvnat}
\bibliography{AsteroidSampleReturnTasker_arxiv}

@article{Jenkins2024,
author = {Jenkins, Laura E. and Lee, Martin R. and Daly, Luke and others},
title = {Winchcombe: An example of rapid terrestrial alteration of a CM chondrite},
journal = {Meteoritics \& Planetary Science},
volume = {59},
number = {5},
pages = {988-1005},
doi = {10.1111/maps.13949},
eprint = {https://onlinelibrary.wiley.com/doi/pdf/10.1111/maps.13949},
year = {2024}
}

@INCOLLECTION{Yoshikawa2015,
       author = {{Yoshikawa}, M. and {Kawaguchi}, J. and {Fujiwara}, A. and {Tsuchiyama}, A.},
        title = "{Hayabusa Sample Return Mission}",
    booktitle = {Asteroids {IV}},
         year = 2015,
       editor = {{Michel}, Patrick and {DeMeo}, Francesca E. and {Bottke}, William F.},
        pages = {397-418},
        publisher = {University of Arizona Press},
          doi = {10.2458/azu_uapress_9780816532131-ch021},
       adsurl = {https://ui.adsabs.harvard.edu/abs/2015aste.book..397Y}
}

@article{Abe2006,
author  = {Abe, M. and Takagi, Y. and Kitazato, K. and others},
title = {Near-Infrared Spectral Results of Asteroid {Itokawa} from the {Hayabusa} Spacecraft},
journal = {Science},
volume = {312},
number = {5778},
pages = {1334--1338},
year = {2006},
doi = {10.1126/science.1125718},
eprint = {https://www.science.org/doi/pdf/10.1126/science.1125718}}

@article{Yurimoto2011,
author = {Hisayoshi Yurimoto  and Ken-ichi Abe  and Masanao Abe  and others},
title = {Oxygen Isotopic Compositions of Asteroidal Materials Returned from {Itokawa} by the {Hayabusa} Mission},
journal = {Science},
volume = {333},
number = {6046},
pages = {1116-1119},
year = {2011},
doi = {10.1126/science.1207776},
eprint = {https://www.science.org/doi/pdf/10.1126/science.1207776}}

@article{Nakamura2011,
year = {2011},
title = {{Itokawa} Dust Particles: a Direct Link Between {S}-Type Asteroids and Ordinary Chondrites},
author = {Nakamura, Tomoki and Noguchi, Takaaki and Tanaka, Masahiko and others},
journal = {Science},
issn = {0036-8075},
doi = {10.1126/science.1207758},
pmid = {21868667},
pages = {1113--1116},
number = {6046},
volume = {333}
}

@article{Fujiwara2006,
year = {2006},
title = {The Rubble-Pile Asteroid {Itokawa} as Observed by {Hayabusa}},
author = {Fujiwara, A. and Kawaguchi, J. and Yeomans, D. K. and others},
journal = {Science},
issn = {0036-8075},
doi = {10.1126/science.1125841},
pmid = {16741107},
pages = {1330--1334},
number = {5778},
volume = {312}
}

@article{Saito2006,
year = {2006},
title = {Detailed Images of Asteroid 25143 {Itokawa} from {Hayabusa}},
author = {Saito, J. and Miyamoto, H. and Nakamura, R. and others},
journal = {Science},
doi = {10.1126/science.1125722},
pmid = {16741110},
pages = {1341--1344},
number = {5778},
volume = {312}
}

@article{Wakita2014,
year = {2014},
title = {{Thermal modeling for a parent body of Itokawa}},
author = {Wakita, S. and Nakamura, T. and Ikeda, T. and Yurimoto, H.},
journal = {Meteoritics \& Planetary Science},
doi = {10.1111/maps.12174},
pages = {228--236},
number = {2},
volume = {49}
}

@article{Michel2009,
year = {2009},
title = {{Itokawa's cratering record as observed by Hayabusa: Implications for its age and collisional history}},
author = {Michel, P. and O'Brien, D.P. and Abe, S. and Hirata, N.},
journal = {Icarus},
doi = {10.1016/j.icarus.2008.04.002},
pages = {503--513},
number = {2},
volume = {200}
}

@article{Michel2006,
year = {2006},
title = {{Dynamical origin of the asteroid (25143) Itokawa: the target of the sample-return Hayabusa space mission}},
author = {Michel, P. and Yoshikawa, M.},
journal = {Astronomy \& Astrophysics},
doi = {10.1051/0004-6361:20054319},
pages = {817--820},
number = {2},
volume = {449}
}

@article{Watanabe2019,
year = {2019},
title = {{Hayabusa2 arrives at the carbonaceous asteroid 162173 Ryugu—A spinning top–shaped rubble pile}},
author = {Watanabe, S and Hirabayashi, M and Hirata, N and others},
journal = {Science},
doi = {10.1126/science.aav8032},
pmid = {30890588},
pages = {268--272},
number = {6437},
volume = {364}
}

@article{Sugita2019,
year = {2019},
title = {{The geomorphology, color, and thermal properties of Ryugu: implications for parent-body processes}},
author = {Sugita, S and Honda, R and Morota, T and others},
journal = {Science},
doi = {10.1126/science.aaw0422},
pmid = {30890587},
pmcid = {PMC7370239},
pages = {252},
number = {6437},
volume = {364}
}

@article{Okada2020,
year = {2020},
title = {{Highly porous nature of a primitive asteroid revealed by thermal imaging}},
author = {Okada, Tatsuaki and Fukuhara, Tetsuya and Tanaka, Satoshi and others},
journal = {Nature Publishing Group},
doi = {10.1038/s41586-020-2102-6},
pages = {1 -- 5},
number = {7800},
volume = {208}
}

@article{Grott2020,
year = {2020},
title = {Macroporosity and Grain Density of Rubble Pile Asteroid (162173) {Ryugu}},
author = {Grott, Matthias and Biele, Jens and Michel, Patrick and others},
journal = {Journal of Geophysical Research: Planets},
doi = {10.1029/2020je006519},
number = {12},
volume = {125}
}

@article{Arakawa2020,
year = {2020},
title = {{An artificial impact on the asteroid (162173) {Ryugu} formed a crater in the gravity-dominated regime}},
author = {Arakawa, M. and Saiki, T. and Wada, K. and others},
journal = {Science},
doi = {10.1126/science.aaz1701},
pmid = {32193363},
pages = {67--71},
number = {6486},
volume = {368}
}

@article{Yokoyama2022,
year = {2022},
title = {{Samples returned from the asteroid Ryugu are similar to Ivuna-type carbonaceous meteorites}},
author = {Yokoyama, Tetsuya and Nagashima, Kazuhide and Nakai, Izumi and others},
journal = {Science},
doi = {10.1126/science.abn7850},
pmid = {35679354},
pages = {eabn7850},
number = {6634},
volume = {379}
}

@article{Paquet2023,
year = {2023},
title = {{Contribution of Ryugu-like material to Earth’s volatile inventory by Cu and Zn isotopic analysis}},
author = {Paquet, Marine and Moynier, Frederic and Yokoyama, Tetsuya and others},
journal = {Nature Astronomy},
doi = {10.1038/s41550-022-01846-1},
pmid = {39776490},
pmcid = {PMC7617279},
pages = {182--189},
number = {2},
volume = {7}
}

@article{Piani2023,
year = {2023},
title = {Hydrogen Isotopic Composition of Hydrous Minerals in Asteroid {Ryugu}},
author = {Piani, Laurette and Nagashima, Kazuhide and Kawasaki, Noriyuki and others},
journal = {The Astrophysical Journal Letters},
doi = {10.3847/2041-8213/acc393},
pages = {L43},
number = {2},
volume = {946}
}

@article{Hu2024,
title = {Pervasive aqueous alteration in the early Solar System revealed by potassium isotopic variations in Ryugu samples and carbonaceous chondrites},
journal = {Icarus},
volume = {409},
pages = {115884},
year = {2024},
doi = {10.1016/j.icarus.2023.115884},
author = {Yan Hu and Frédéric Moynier and Wei Dai and others}
}

@article{Nakamura2022,
year = {2022},
title = {{Formation and evolution of carbonaceous asteroid Ryugu: direct evidence from returned samples}},
author = {Nakamura, T and Matsumoto, M and Amano, K and others},
journal = {Science},
doi = {10.1126/science.abn8671},
pmid = {36137011},
pages = {eabn8671},
number = {6634},
volume = {379}
}

@article{Tachibana2022,
year = {2022},
title = {{Pebbles and sand on asteroid (162173) Ryugu: In situ observation and particles returned to Earth}},
author = {Tachibana, S and Sawada, H and Okazaki, R and others},
journal = {Science},
doi = {10.1126/science.abj8624},
pmid = {35143255},
pages = {1011--1016},
number = {6584},
volume = {375}
}

@article{NakamuraEizo2022,
year = {2022},
title = {{On the origin and evolution of the asteroid Ryugu: A comprehensive geochemical perspective}},
author = {Nakamura, Eizo and Kobayashi, Katsura and Tanaka, Ryoji and others},
journal = {Proceedings of the Japan Academy, Series B},
doi = {10.2183/pjab.98.015},
pmid = {35691845},
pmcid = {PMC9246647},
pages = {227--282},
number = {6},
volume = {98}
}

@article{Kitazato2019,
year = {2019},
title = {{The surface composition of asteroid 162173 Ryugu from Hayabusa2 near-infrared spectroscopy}},
author = {Kitazato, K and Milliken, R E and Iwata, T and others},
journal = {Science},
doi = {10.1126/science.aav7432},
pmid = {30890589},
pages = {272--275},
number = {6437},
volume = {364}
}

@article{Naraoka2023,
year = {2023},
title = {{Soluble organic molecules in samples of the carbonaceous asteroid (162173) Ryugu}},
author = {Naraoka, Hiroshi and Takano, Yoshinori and Dworkin, Jason P. and others},
journal = {Science},
doi = {10.1126/science.abn9033},
pmid = {36821691},
pages = {eabn9033},
number = {6634},
volume = {379}
}

@article{Oba2023,
year = {2023},
title = {{Uracil in the carbonaceous asteroid (162173) Ryugu}},
author = {Oba, Yasuhiro and Koga, Toshiki and Takano, Yoshinori and others},
journal = {Nature Communications},
doi = {10.1038/s41467-023-36904-3},
pmid = {36944653},
pmcid = {PMC10030641},
pages = {1292},
number = {1},
volume = {14}
}

@article{Barosch2022,
year = {2022},
title = {Presolar Stardust in Asteroid {Ryugu}},
author = {Barosch, Jens and Nittler, Larry R. and Wang, Jianhua and others},
journal = {The Astrophysical Journal Letters},
doi = {10.3847/2041-8213/ac83bd},
eprint = {2208.07976},
pages = {L3},
number = {1},
volume = {935}
}

@article{Zeichner2023,
year = {2023},
title = {{Polycyclic aromatic hydrocarbons in samples of Ryugu formed in the interstellar medium}},
author = {Zeichner, Sarah S. and Aponte, José C. and Bhattacharjee, Surjyendu and others},
journal = {Science},
doi = {10.1126/science.adg6304},
pmid = {38127762},
pages = {1411--1416},
number = {6677},
volume = {382}
}

@article{Morota2020,
year = {2020},
title = {{Sample collection from asteroid (162173) Ryugu by Hayabusa2: implications for surface evolution}},
author = {Morota, T and Sugita, S and Cho, Y and others},
journal = {Science},
doi = {10.1126/science.aaz6306},
pmid = {32381723},
pages = {654--659},
number = {6491},
volume = {368}
}

@article{OkazakiGasSample2022,
year = {2022},
title = {{First asteroid gas sample delivered by the Hayabusa2 mission: A treasure box from Ryugu}},
author = {Okazaki, Ryuji and Miura, Yayoi N. and Takano, Yoshinori and others},
journal = {Science Advances},
doi = {10.1126/sciadv.abo7239},
pmid = {36264781},
pmcid = {PMC11627213},
pages = {eabo7239},
number = {46},
volume = {8}
}

@article{Okazaki2022,
year = {2022},
title = {{Noble gases and nitrogen in samples of asteroid Ryugu record its volatile sources and recent surface evolution}},
author = {Okazaki, Ryuji and Marty, Bernard and Busemann, Henner and others},
journal = {Science},
doi = {10.1126/science.abo0431},
pmid = {36264828},
pages = {eabo0431},
number = {6634},
volume = {379}
}

@article{Noguchi2023,
year = {2023},
title = {{A dehydrated space-weathered skin cloaking the hydrated interior of Ryugu}},
author = {Noguchi, Takaaki and Matsumoto, Toru and Miyake, Akira and others},
journal = {Nature Astronomy},
doi = {10.1038/s41550-022-01841-6},
pmid = {36845884},
pmcid = {PMC9943745},
pages = {170--181},
number = {2},
volume = {7}
}

@INCOLLECTION{Weisberg2006,
author = {{Weisberg}, M.~K. and {McCoy}, T.~J. and {Krot}, A.~N.},
title = "{Systematics and evaluation of meteorite classification}",
booktitle = {Meteorites and the early Solar System II},
year = {2006},
month = {07},
editor = {{Lauretta}, Dante S. and {McSween}, Harry Y.},
publisher = {University of Arizona Press},
pages = {19},
isbn = {9780816525621},
doi = {10.2307/j.ctv1v7zdmm.8}
}

@article{Kaasalainen2003,
year = {2003},
title = {{CCD photometry and model of MUSES-C target (25143) 1998 SF36}},
author = {Kaasalainen, M. and Kwiatkowski, T. and Abe, M. and Piironen, J. and others},
journal = {Astronomy \& Astrophysics},
doi = {10.1051/0004-6361:20030819},
pages = {L29--L32},
number = {3},
volume = {405}
}

@article{Ostro2005,
year = {2005},
title = {{Radar observations of Itokawa in 2004 and improved shape estimation}},
author = {Ostro, Steven J. and Benner, Lance A. M. and Magri, Christopher and others},
journal = {Meteoritics \& Planetary Science},
doi = {10.1111/j.1945-5100.2005.tb00131.x},
pages = {1563--1574},
number = {11},
volume = {40}
}

@article{DeMeo2009,
year = {2009},
title = {{An extension of the Bus asteroid taxonomy into the near-infrared}},
author = {DeMeo, Francesca E. and Binzel, Richard P. and Slivan, Stephen M. and Bus, Schelte J.},
journal = {Icarus},
doi = {10.1016/j.icarus.2009.02.005},
pages = {160--180},
number = {1},
volume = {202}
}

@INCOLLECTION{Cheng2002,
       author = {{Cheng}, A.~F.},
        title = "{Near-Earth asteroid rendezvous: mission summary}",
    booktitle = {Asteroids III},
         year = 2002,
       editor = {{Bottke}, Jr., W.~F. and {Cellino}, A. and {Paolicchi}, P. and {Binzel}, R.~P.},
        pages = {351-366},
        doi = {10.2307/j.ctv1v7zdn4.29},
    publisher = {University of Arizona Press},
}

@article{Yada2014,
year = {2014},
title = {{Hayabusa‐returned sample curation in the Planetary Material Sample Curation Facility of JAXA}},
author = {Yada, Toru and Fujimura, Akio and Abe, Masanao and others},
journal = {Meteoritics \& Planetary Science},
issn = {1086-9379},
doi = {10.1111/maps.12027},
pages = {135--153},
number = {2},
volume = {49},
}

@article{Noguchi2014,
year = {2014},
title = {{Space weathered rims found on the surfaces of the Itokawa dust particles}},
author = {Noguchi, Takaaki and Kimura, Makoto and Hashimoto, Takahito and others},
journal = {Meteoritics \& Planetary Science},
doi = {10.1111/maps.12111},
pages = {188--214},
number = {2},
volume = {49},
}

@article{Bottke2025,
year = {2025},
title = {Surface Ages for the Sample Return Asteroids {Bennu}, {Ryugu}, and {Itokawa}},
author = {Bottke, William F. and Meyer, Alex J. and Vokrouhlický, David and others},
journal = {The Planetary Science Journal},
doi = {10.3847/psj/add46a},
pages = {150},
number = {6},
volume = {6},
}

@article{Tachibana2014,
year = {2014},
title = {{Hayabusa2: Scientific importance of samples returned from C-type near-Earth asteroid (162173) 1999 JU3}},
journal = {Geochemical Journal},
author = {Tachibana, S. and Abe, M. and Arakawa, M. and others},
doi = {10.2343/geochemj.2.0350},
pages = {571--587},
number = {6},
volume = {48},
}

@article{Sawada2017,
year = {2017},
title = {Hayabusa2 Sampler: collection of Asteroidal Surface Material},
author = {Sawada, Hirotaka and Okazaki, Ryuji and Tachibana, Shogo and others},
journal = {Space Science Reviews},
doi = {10.1007/s11214-017-0338-8},
pages = {81--106},
number = {1-4},
volume = {208},
}

@article{Sugawara2024,
year = {2024},
title = {{Update on the 53Mn-53Cr ages of dolomite in the Ivuna CI chondrite and asteroid Ryugu sample}},
author = {Sugawara, Shingo and Fujiya, Wataru and Kawasaki, Noriyuki and others},
journal = {Geochimica et Cosmochimica Acta},
doi = {10.1016/j.gca.2024.08.013},
pages = {40--50},
volume = {382},
}

@article{Kita2024,
year = {2024},
title = {{Disequilibrium oxygen isotope distribution among aqueously altered minerals in Ryugu asteroid returned samples}},
author = {Kita, Noriko T. and Kitajima, Kouki and Nagashima, Kazuhide and others},
journal = {Meteoritics \& Planetary Science},
doi = {10.1111/maps.14163},
pages = {2097--2116},
number = {8},
volume = {59},
}

@article{Yabuta2023,
year = {2023},
title = {{Macromolecular organic matter in samples of the asteroid (162173) Ryugu}},
author = {Yabuta, Hikaru and Cody, George D. and Engrand, Cécile and others},
journal = {Science},
doi = {10.1126/science.abn9057},
pmid = {36821663},
pages = {eabn9057},
number = {6634},
volume = {379},
}

@article{Righter2023,
year = {2023},
title = {{Curation planning and facilities for asteroid Bennu samples returned by the OSIRIS‐REx mission}},
author = {Righter, K. and Lunning, N. G. and Nakamura‐Messenger, K. and others},
journal = {Meteoritics \& Planetary Science},
doi = {10.1111/maps.13973},
pages = {572--590},
number = {4},
volume = {58},
}

@article{Lauretta2019,
year = {2019},
title = {{The unexpected surface of asteroid (101955) Bennu}},
author = {Lauretta, D. S. and DellaGiustina, D. N. and Bennett, C. A. and others},
journal = {Nature},
doi = {10.1038/s41586-019-1033-6},
pmid = {30890786},
pmcid = {PMC6557581},
pages = {55--60},
number = {7750},
volume = {568},
}

@article{DellaGiustina2019,
year = {2019},
title = {{Properties of rubble-pile asteroid (101955) Bennu from OSIRIS-REx imaging and thermal analysis}},
author = {DellaGiustina, D. N. and Emery, J. P. and Golish, D. R. and others},
journal = {Nature Astronomy},
doi = {10.1038/s41550-019-0731-1},
pages = {341--351},
number = {4},
volume = {3},
}

@article{Rozitis2020,
year = {2020},
title = {{Asteroid (101955) Bennu’s weak boulders and thermally anomalous equator}},
author = {Rozitis, B. and Ryan, A. J. and Emery, J. P. and others},
journal = {Science Advances},
doi = {10.1126/sciadv.abc3699},
pmid = {33033037},
pmcid = {PMC7544501},
pages = {eabc3699},
number = {41},
volume = {6}
}

@article{Daly2020,
year = {2020},
title = {{Hemispherical differences in the shape and topography of asteroid (101955) Bennu}},
author = {Daly, M. G. and Barnouin, O. S. and Seabrook, J. A. and others},
journal = {Science Advances},
doi = {10.1126/sciadv.abd3649},
pmid = {33033038},
pmcid = {PMC7544500},
pages = {eabd3649},
number = {41},
volume = {6},
}

@article{Sheeres2020,
year = {2020},
title = {{Heterogeneous mass distribution of the rubble-pile asteroid (101955) Bennu}},
author = {Scheeres, D. J. and French, A. S. and Tricarico, P. and others},
journal = {Science Advances},
doi = {10.1126/sciadv.abc3350},
pmid = {33033036},
pmcid = {PMC7544499},
pages = {eabc3350},
number = {41},
volume = {6}
}

@article{Scheeres2019,
year = {2019},
title = {{The dynamic geophysical environment of (101955) Bennu based on OSIRIS-REx measurements}},
author = {Scheeres, D. J. and McMahon, J. W. and French, A. S. and others},
journal = {Nature Astronomy},
doi = {10.1038/s41550-019-0721-3},
pmid = {32601603},
pmcid = {PMC7323631},
pages = {352--361},
number = {4},
volume = {3}
}

@article{LaurettaConnolly2024,
year = {2024},
title = {Asteroid (101955) {Bennu} in the laboratory: properties of the sample collected by {OSIRIS‐REx}},
author = {Lauretta, Dante S. and Connolly, Harold C. and others},
journal = {Meteoritics \& Planetary Science},
doi = {10.1111/maps.14227},
pages = {2453--2486},
number = {9},
volume = {59}
}

@article{ConnollyLauretta2025,
year = {2025},
title = {{An overview of the petrography and petrology of particles from aggregate sample from asteroid Bennu}},
author = {Connolly, Harold C. and Lauretta, Dante S. and others},
journal = {Meteoritics \& Planetary Science},
doi = {10.1111/maps.14335},
pages = {979--996},
number = {5},
volume = {60}
}

@article{Ye2019,
doi = {10.3847/2515-5172/ab12e7},
year = {2019},
month = {mar},
publisher = {The American Astronomical Society},
volume = {3},
number = {3},
pages = {56},
author = {Ye, Quanzhi},
title = {Prediction of Meteor Activities from (101955) {Bennu}},
journal = {Research Notes of the AAS}
}

@article{Chesley2014,
title = {Orbit and bulk density of the {OSIRIS-REx} target Asteroid (101955) {Bennu}},
journal = {Icarus},
volume = {235},
pages = {5-22},
year = {2014},
doi = {10.1016/j.icarus.2014.02.020},
author = {Steven R. Chesley and Davide Farnocchia and Michael C. Nolan and others}
}

@article{Farnocchia2021,
year = {2021},
title = {{Ephemeris and hazard assessment for near-Earth asteroid (101955) Bennu based on OSIRIS-REx data}},
author = {Farnocchia, Davide and Chesley, Steven R. and Takahashi, Yu and others},
journal = {Icarus},
issn = {0019-1035},
doi = {10.1016/j.icarus.2021.114594},
pages = {114594},
volume = {369}
}

@article{Hung2023,
doi = {10.3847/PSJ/ad0226},
year = {2023},
publisher = {The American Astronomical Society},
volume = {4},
number = {11},
pages = {215},
author = {Hung, Denise and Tholen, David J. and Farnocchia, Davide and Spoto, Federica},
title = {Detectability of the {Yarkovsky} Effect in the {Main Belt}},
journal = {The Planetary Science Journal}
}

@article{Hamilton2019,
year = {2019},
title = {{Evidence for widespread hydrated minerals on asteroid (101955) Bennu}},
author = {Hamilton, V. E. and Simon, A. A. and Christensen, P. R. and others},
journal = {Nature Astronomy},
doi = {10.1038/s41550-019-0722-2},
pmid = {31360777},
pmcid = {PMC6662227},
pages = {332--340},
number = {4},
volume = {3}
}

@article{Barnes2025,
year = {2025},
title = {{The variety and origin of materials accreted by Bennu’s parent asteroid}},
author = {Barnes, J. J. and Nguyen, A. N. and Abernethy, F. A. J. and others},
journal = {Nature Astronomy},
doi = {10.1038/s41550-025-02631-6},
pages = {1--18}
}

@article{Zega2025,
year = {2025},
title = {{Mineralogical evidence for hydrothermal alteration of Bennu samples}},
author = {Zega, T. J. and McCoy, T. J. and Russell, S. S. and others},
journal = {Nature Geoscience},
doi = {10.1038/s41561-025-01741-0},
pages = {1--8}
}

@article{McCoy2025,
year = {2025},
title = {{An evaporite sequence from ancient brine recorded in Bennu samples}},
author = {McCoy, T. J. and Russell, S. S. and Zega, T. J. and others},
journal = {Nature},
doi = {10.1038/s41586-024-08495-6},
pmid = {39880992},
pmcid = {PMC11779627},
pages = {1072--1077},
number = {8048},
volume = {637}
}

@article{Sekine2025,
year = {2025},
title = {{Asteroid Bennu contains salts from ancient brine}},
author = {Sekine, Yasuhito},
journal = {Nature},
doi = {10.1038/d41586-025-00084-5},
pmid = {39880989},
pages = {1056--1058},
number = {8048},
volume = {637}
}

@article{Glavin2025,
year = {2025},
title = {{Abundant ammonia and nitrogen-rich soluble organic matter in samples from asteroid (101955) Bennu}},
author = {Glavin, Daniel P. and Dworkin, Jason P. and Alexander, Conel M. O’D. and others},
journal = {Nature Astronomy},
doi = {10.1038/s41550-024-02472-9},
pmid = {39990238},
pmcid = {PMC11842271},
pages = {199--210},
number = {2},
volume = {9}
}

@article{Matsumoto2024,
year = {2024},
title = {{Sodium carbonates on Ryugu as evidence of highly saline water in the outer Solar System}},
author = {Matsumoto, Toru and Noguchi, Takaaki and Miyake, Akira and others},
journal = {Nature Astronomy},
doi = {10.1038/s41550-024-02418-1},
pages = {1536--1543},
number = {12},
volume = {8}
}

@article{Keller2025,
year = {2025},
title = {{Space weathering effects in Bennu asteroid samples}},
author = {Keller, L. P. and Thompson, M. S. and Seifert, L. B. and others},
journal = {Nature Geoscience},
issn = {1752-0894},
doi = {10.1038/s41561-025-01745-w},
pages = {1--7}
}

@article{Arredondo2025,
year = {2025},
title = {{JWST Spectroscopy of (142) Polana: Connection to NEAs (101955) Bennu and (162173) Ryugu}},
author = {Arredondo, Anicia and Becker, Tracy M. and McAdam, Maggie M. and others},
journal = {The Planetary Science Journal},
doi = {10.3847/psj/ade395},
pages = {195},
number = {8},
volume = {6}
}

@article{Noguchi2025,
year = {2025},
title = {{Clues from asteroids about Earth’s volatiles}},
author = {Noguchi, Takaaki},
journal = {Nature Geoscience},
doi = {10.1038/s41561-025-01775-4},
pages = {812--814},
number = {9},
volume = {18}
}

@article{Nakato2022,
year = {2022},
title = {{Ryugu particles found outside the Hayabusa2 sample container}},
author = {Nakato, Aiko and Inada, Shiori and Furuya, Shizuho and others},
journal = {Geochemical Journal},
doi = {10.2343/geochemj.gj22017},
pages = {197--222},
number = {6},
volume = {56},
}

@article{Noguchi2024,
year = {2024},
title = {{Mineralogy and petrology of fine‐grained samples recovered from the asteroid (162173) Ryugu}},
author = {Noguchi, Takaaki and Matsumoto, Toru and Miyake, Akira and others},
journal = {Meteoritics \& Planetary Science},
doi = {10.1111/maps.14093},
pages = {1877--1906},
number = {8},
volume = {59}
}

@article{Koga2026,
year = {2026},
title = {{A complete set of canonical nucleobases in the carbonaceous asteroid (162173) Ryugu}},
author = {Koga, Toshiki and Oba, Yasuhiro and Takano, Yoshinori and others},
journal = {Nature Astronomy},
doi = {10.1038/s41550-026-02791-z},
pages = {1--9}
}

\end{document}